# Physical Interpretations of Nilpotent Quantum Mechanics

## Peter Rowlands

*Department of Physics, University of Liverpool, Oliver Lodge Laboratory, Oxford Street, Liverpool, L69 7ZE, UK. e-mail p.rowlands@liverpool.ac.uk*

*Abstract.* Nilpotent quantum mechanics provides a powerful method of making efficient calculations. More importantly, however, it provides insights into a number of fundamental physical problems through its use of a dual vector space and its explicit construction of vacuum. Physical interpretation of the nilpotent formalism is discussed with respect to boson and baryon structures, the mass-gap problem, *zitterbewgung*, Berry phase, renormalization, and related issues.

**Keywords:** nilpotent; fermion; duality, zero totality; vacuum, vector space

## 1 Two vector spaces

Nilpotent quantum mechanics [1-12] has its origin in the conventional form of the Dirac equation, structured in terms of the 4 × 4 $\gamma$ matrices:

$$\left(\gamma^\mu \partial_\mu + im\right)\psi = \left(\gamma^0 \frac{\partial}{\partial t} + \gamma^1 \frac{\partial}{\partial x} + \gamma^2 \frac{\partial}{\partial y} + \gamma^3 \frac{\partial}{\partial y} + im\right)\psi = 0 \qquad (1)$$

The $\gamma$ matrices define a 64-part algebra, incorporating all possible algebraic products of the five matrices: $\gamma^0$, $\gamma^1$, $\gamma^2$, $\gamma^3$, $\gamma^5$. The three terms $\gamma^1$, $\gamma^2$, $\gamma^3$ are taken as mutually orthogonal components of an object with vector properties (**γ**), which, additionally, incorporates the property of half-integral spin. The units of the underlying vector algebra can be represented by the 2 × 2 unit Pauli matrices, $\boldsymbol{\sigma}^1$, $\boldsymbol{\sigma}^2$, $\boldsymbol{\sigma}^3$, with multiplication properties

$$(\boldsymbol{\sigma}^1)^2 = (\boldsymbol{\sigma}^2)^2 = (\boldsymbol{\sigma}^3)^2 = \mathbf{I} \text{ (unit matrix)};$$
$$\boldsymbol{\sigma}^1\boldsymbol{\sigma}^2 = -\boldsymbol{\sigma}^2\boldsymbol{\sigma}^1 = i\boldsymbol{\sigma}^3;\ \boldsymbol{\sigma}^2\boldsymbol{\sigma}^3 = -\boldsymbol{\sigma}^3\boldsymbol{\sigma}^2 = i\boldsymbol{\sigma}^1;\ \boldsymbol{\sigma}^3\boldsymbol{\sigma}^1 = -\boldsymbol{\sigma}^1\boldsymbol{\sigma}^3 = i\boldsymbol{\sigma}^2,\ \boldsymbol{\sigma}^1\boldsymbol{\sigma}^2\boldsymbol{\sigma}^3 = i\mathbf{I}. \qquad (2)$$

They are isomorphic to the *multivariate vectors*, **i**, **j**, **k**, defined by Hestenes [13], with multiplication rules

$$\mathbf{i}^2 = \mathbf{j}^2 = \mathbf{k}^2 = 1 \text{ (unit scalar)}$$
$$\mathbf{ij} = -\mathbf{ji} = i\mathbf{k};\ \mathbf{jk} = -\mathbf{kj} = i\mathbf{i};\ \mathbf{ki} = -\mathbf{ik} = i\mathbf{j}. \qquad (3)$$

which again are isomorphic to the complexified quaternions (*i**i***), (*i**j***), (*i**k***), defined by the rules:

$$(i\mathbf{i})^2 = (i\mathbf{j})^2 = (i\mathbf{k})^2 = 1 \text{ (unit scalar)};$$
$$(i\mathbf{i})(i\mathbf{j}) = -(i\mathbf{j})(i\mathbf{i}) = i(i\mathbf{k});\ (i\mathbf{j})(i\mathbf{k}) = -(i\mathbf{k})(i\mathbf{j}) = i(i\mathbf{i});\ (i\mathbf{k})(i\mathbf{i}) = -(i\mathbf{i})(i\mathbf{k}) = i(i\mathbf{j}). \qquad (4)$$

The multivariate vectors are also defined as the real Pauli Clifford algebra of Euclidean 3 space, of dimension $2^3 = 8$. They create the complete algebra of 3-dimensional Euclidean space, incorporating also the scalar (1), pseudoscalar (*i*), vector (**i**, **j**, **k**), and pseudovector (*i***i**, *i***j**, *i***k**) units needed to specify all aspects of spatial definition, including volumes and areas. They are distinguished from ordinary vectors only in defining a full or algebraic product between vectors **a** and **b**, which combines scalar and vector products:

$$\mathbf{ab} = \mathbf{a}.\mathbf{b} + i\, \mathbf{a} \times \mathbf{b}. \tag{5}$$

The extra cross-product term has long been established as the ultimate source of fermionic spin [13].

The 2 × 2 Pauli matrices, $\sigma^1, \sigma^2, \sigma^3$, are clearly an intrinsic structural component of the 4 × 4 Dirac or $\gamma$ matrices. In fact, their complete specification requires only another set of Pauli matrices, say, $\Sigma^1, \Sigma^2, \Sigma^3$, to which they are commutative. The result of all possible *commutative* multiplications of the six units $\sigma^1, \sigma^2, \sigma^3, \Sigma^1, \Sigma^2, \Sigma^3$, including the unit 2 × 2 matrices and complexified unit matrices which they generate (e.g. as $\sigma^1\sigma^1$ and $\sigma^1\sigma^2\sigma^3$ – we will use, for convenience, the ones, say **I** and *i***I**, which are generated by the $\sigma$ set), are a group of 64 4 × 4 matrices.

Now, a group of order 64, is minimally generated by five elements. Though this can be done in many ways, all sets of five generating elements will have exactly the same structure, typically given by:

$$\Sigma^1\mathbf{I}, i\sigma^1\Sigma^3, i\sigma^2\Sigma^3, i\sigma^3\Sigma^3, i\Sigma^2\mathbf{I} \tag{6}$$

or

$$\Sigma^3\mathbf{I}, i\sigma^1\Sigma^1, i\sigma^2\Sigma^1, i\sigma^3\Sigma^1, i\Sigma^2\mathbf{I} \tag{7}$$

which correspond respectively to $\gamma^0, \gamma^1, \gamma^2, \gamma^3, \gamma^5$. The allowed variations include interchanging the three $\Sigma$ units with each other (as between (6) and (7)); similarly interchanging the three $\sigma$ units; and interchanging the two commutative sets (together with the unit matrix). In addition, any two of $\sigma^1, \sigma^2, \sigma^3$ or $\Sigma^1, \Sigma^2, \Sigma^3$ can simultaneously change sign. The commutative multiplication ensures that all the resulting matrices are 4 × 4, and all possible versions of the 4 × 4 Dirac matrices can be derived in this manner.

From a physical point of view, this representation has two interlocking 3-dimensional structures, one in which the rotational symmetry of the components preserved, and the other in which the symmetry is broken. This becomes more readily apparent when we switch the representation from matrices to various forms of geometrical algebra. Here, we can replace $\sigma^1, \sigma^2, \sigma^3$ and $\Sigma^1, \Sigma^2, \Sigma^3$ by two commutative systems of multivariate vectors, two commutative real Pauli Clifford algebras of Euclidean 3 space, or two sets of complexified quaternions. We can even commutatively combine, without loss of generality, a system of quaternions (*i, j, k*) with a set of multivariate vectors (**i**, **j**, **k**) or units of geometrical algebra (**e₁**, **e₂**, **e₃**).

In previous work, we have used the algebra formed by the commutative combination of quaternions and multivariate vectors in + and – values of eight basic units (1, *i*, *i*, *j*, *k*, **i**, **j**, **k**). This is interestingly close to Penrose's twistor structure [14] in having 4 'real' parts (norm 1) and 4 'imaginary' parts (norm −1), but it also has additional structure, and this turns out to be crucial in separating out two sets of 3-dimensional objects and two full vector spaces. The vector units (**i**, **j**, **k**) will be mostly subsumed into a single momentum operator, so the problem of confusion with the quaternion units (*i*, *j*, *k*) should not arise. However, the signs of the quaternion units will, of course, be arbitrary, so we may choose them purely for convenience, and replace the $\gamma$ representations in (7) by, say,

$$-i\mathbf{k}, -i\boldsymbol{\sigma}^1, -i\boldsymbol{\sigma}^2, -i\boldsymbol{\sigma}^3, i\mathbf{j} \tag{8}$$

or

$$-i\mathbf{k}, -i\mathbf{i}, -i\mathbf{j}, -i\mathbf{k}, i\mathbf{j} \tag{9}$$

The standard form of the Dirac equation, represented by (1), does not incorporate $\gamma^5$, although this is a fundamental component of the algebra. To obtain a more symmetric structure, we premultiply by $-i\gamma^5$:

$$-i\gamma^5\left(\gamma^\mu \partial_\mu + im\right)\psi = -i\gamma^5\left(\gamma^0 \frac{\partial}{\partial t} + \gamma^1 \frac{\partial}{\partial x} + \gamma^2 \frac{\partial}{\partial y} + \gamma^3 \frac{\partial}{\partial y} + im\right)\psi = 0. \tag{10}$$

With the representations of the $\gamma$ matrices defined in (6), we can write (10) in the form

$$\boldsymbol{\Sigma}^2\mathbf{I}\left(\boldsymbol{\Sigma}^1\mathbf{I}\frac{\partial}{\partial t} + i\boldsymbol{\Sigma}^3\boldsymbol{\sigma}^1\frac{\partial}{\partial x} + i\boldsymbol{\Sigma}^3\boldsymbol{\sigma}^2\frac{\partial}{\partial y} + i\boldsymbol{\Sigma}^3\boldsymbol{\sigma}^3\frac{\partial}{\partial y} + im\right)\psi = 0 \tag{11}$$

or

$$\left(-i\boldsymbol{\Sigma}^3\mathbf{I}\frac{\partial}{\partial t} - \boldsymbol{\Sigma}^1\boldsymbol{\sigma}^1\frac{\partial}{\partial x} - \boldsymbol{\Sigma}^1\boldsymbol{\sigma}^2\frac{\partial}{\partial y} - \boldsymbol{\Sigma}^1\boldsymbol{\sigma}^3\frac{\partial}{\partial y} + i\boldsymbol{\Sigma}^2\mathbf{I}m\right)\psi = 0. \tag{12}$$

**2 The nilpotent Dirac equation**

We have already established that the labellings of the $\Sigma$ terms are intrinsically arbitrary, and any two can change sign, so (12) differs from (1) only in having $i\Sigma^2\mathbf{I}$ or $\gamma^5$ rather than *i* as coefficient of *m*. However, there is a significant *physical* difference in that the two vector spaces, $\boldsymbol{\sigma}^1, \boldsymbol{\sigma}^2, \boldsymbol{\sigma}^3$ and $\boldsymbol{\Sigma}^1, \boldsymbol{\Sigma}^2, \boldsymbol{\Sigma}^3$, are now fully incorporated into the equation. It also becomes readily apparent that the $\gamma$ matrices, or their algebraic equivalents, are to be found within the wavefunction $\psi$, as well as in the operator. Although we could conceivably do this using matrices, it is much more convenient to use the algebraic equivalents, defined in (9). We then obtain

$$\left(-\mathbf{k}\frac{\partial}{\partial t} - i\mathbf{i}\frac{\partial}{\partial x} - i\mathbf{i}\mathbf{j}\frac{\partial}{\partial y} - i\mathbf{i}\mathbf{k}\frac{\partial}{\partial y} + \mathbf{j}m\right)\psi = \left(-\mathbf{k}\frac{\partial}{\partial t} - i\mathbf{i}\nabla + \mathbf{j}m\right)\psi = 0. \tag{13}$$

As soon as we insert a plane wave solution, for a free particle, into (13), say

$$\psi = A\, e^{-i(Et - \mathbf{p}\cdot\mathbf{r})}, \tag{14}$$

we obtain

$$\left(-\bm{k}\frac{\partial}{\partial t} - \bm{ii}\nabla + \bm{j}m\right)Ae^{-i(Et-\mathbf{p}.\mathbf{r})} = \left(i\bm{k}E + i\bm{ip} + \bm{j}m\right)Ae^{-i(Et-\mathbf{p}.\mathbf{r})} = 0. \qquad (15)$$

The condition for this to be true is that $A$ becomes identical to ($i\bm{k}E + i\bm{ip} + \bm{j}m$) or a scalar multiple of it, which means that both ($i\bm{k}E + i\bm{ip} + \bm{j}m$) and $\psi$ are *nilpotents* or square roots of 0. (Here, we need to specify that the multivariate properties of **p** allow us to use the 'spin' terms **p** and $\nabla$ instead of the 'helicity' terms $\sigma.\mathbf{p}$ and $\sigma.\nabla$, where $\sigma$ is a unit pseudovector of magnitude $-1$, in a nilpotent structure, since $(\sigma.\mathbf{p})^2 = \mathbf{pp} = p^2$.) Now $\psi$ is not, of course, a single term but rather a 4-component spinor, incorporating fermion / antifermion and spin up / down states. However, this is easily accommodated by transforming ($i\bm{k}E + i\bm{ip} + \bm{j}m$) into a column vector with four sign combinations of the $E$ and **p** terms, which may be written in abbreviated form as ($\pm i\bm{k}E \pm i\bm{ip} + \bm{j}m$). In principle, the same four sign options should also apply to the phase factor, $e^{-i(Et - \mathbf{p}.\mathbf{r})}$, but an alternative representation allows us to apply the variations instead to the differential operator, if this is now restructured as a 4-component row vector. This means that our final equation becomes

$$\left(\mp\bm{k}\frac{\partial}{\partial t} \mp \bm{ii}\nabla + \bm{j}m\right)\left(\pm i\bm{k}E \pm i\bm{ip} + \bm{j}m\right)e^{-i(Et-\mathbf{p}.\mathbf{r})} = 0 \qquad (16)$$

where both operator and amplitude are 4-component spinors, and the Feynman principle of particles having negative energy states also having reversed time direction becomes an immediate consequence. This is the Dirac equation for a free fermion with nilpotent wavefunction.

Though conventional relativistic quantum mechanics has been assumed to require idempotent, rather than nilpotent wavefunctions, i.e. ones that square to themselves ($\psi\psi \to \psi$) rather than to zero, exactly the same equation can be read as either idempotent or nilpotent, by a simple redistribution of a single algebraic unit between the sections of the equation defined as operator and wavefunction:

$$[(\mp \bm{k}\partial/\partial t \mp \bm{ii}\nabla + \bm{j}m)\bm{j}]\ [\bm{j}(\pm i\bm{k}E \pm i\bm{ip} + \bm{j}m)\ e^{-i(Et - \mathbf{p}.\mathbf{r})}] = 0.$$
  *operator*      *idempotent wavefunction*

$$[(\mp \bm{k}\partial/\partial t \mp \bm{ii}\nabla + \bm{j}m)\bm{jj}]\ [(\pm i\bm{k}E \pm i\bm{ip} + \bm{j}m)\ e^{-i(Et - \mathbf{p}.\mathbf{r})}] = 0.$$
  *operator*      *nilpotent wavefunction*

Mathematically, the two interpretations are equivalent, and the relativistic wavefunction has both properties. The idempotent aspect is important in establishing the 4-component wavefunction as a spinor, though this also carries over from the conventional version of relativistic quantum mechanics. A column vector of the form $\bm{j}(\pm i\bm{k}E \pm i\bm{ip} + \bm{j}m)$ has four idempotent terms which sum up to unity after normalization and, when combined with the appropriate quaternionic coefficients identifying their positions in the column, produce zero products between any two terms as required (in principle, demonstrating their intrinsic nilpotency). The nilpotent vector ($\pm ikE \pm i\mathbf{p} + jm$) also has this intrinsic idempotent property, as the initial factor $j$ may be subsumed into the quaternion coefficients. The idempotent property, as we will show, has, in addition, a special physical significance, besides helping to define spinor characteristics, but the nilpotent property has a richer structure. The choice of quantum formalisms is not a neutral one in determining the characteristics of physical systems. Nilpotency, as we will show, is a statement of a *physical* principle, rather than a purely mathematical operation.

The nilpotent formalism gains a greatly increased calculating power through its expression in terms of a single phase factor. Even more significant is the physical information incorporated within the nilpotent structure. In principle, a particle with a nilpotent wavefunction, say $\psi_1$, becomes automatically Pauli exclusive, by zeroing the combination state with an identical particle $\psi_1 \psi_1$. The important extension occurs, however, when the fermions are no longer free, but subject to forces from other fermions, as in all cases of Pauli exclusion so far observed. The nilpotent formalism can accommodate this without difficulty by redefining the operators $E$ and $\mathbf{p}$ to incorporate field terms or covariant derivatives, so that $E$ now becomes, say, $i\partial / \partial t + e\phi + \ldots$, and $\mathbf{p}$ becomes, say, $-i\nabla + e\mathbf{A} + \ldots$. The eigenvalues $E$ and $\mathbf{p}$ will then represent more complicated expressions resulting from the presence of the additional terms in the operator, and the phase factor will no longer be the $e^{-i(Et - \mathbf{p}\cdot\mathbf{r})}$ for the free particle, but the properties of the system will still be determined by the need to maintain Pauli exclusion for all fermions, whether free or interacting. The same will also be true if the external field terms are defined by expectation values, as with the Lamb shift, or in terms of quantum fields.

The reduction to a single phase factor and the extra constraint of nilpotency mean that much of the formal apparatus of relativistic quantum mechanics becomes redundant. Writing the operator in the form ($\pm ikE \pm i\mathbf{p} + jm$), with $E$ and $\mathbf{p}$ defined as generic terms involving differentials and associated potentials, specifies the entire quantum mechanics of the system, with the wavefunction and even the equation losing their status as independent sources of information. The operator alone uniquely determines the phase factor which is needed to create a nilpotent amplitude, and this requires only a secondary functional equation: (operator acting on phase factor)$^2$ = 0. Even the spinor representation loses its fundamental status, as the first of the four terms, say ($ikE + i\mathbf{p} + jm$), uniquely specifies the remaining three by automatic sign variation, and it will often be convenient to specify the operator in this abbreviated form.

## 3 The 4-component spinor

Conventionally, the Dirac equation produces a wavefunction which is a *spinor*, with four components, which may be structured as a column vector, representing the four combinations of particle and antiparticle, and spin up and spin down. Taking $\pm E$ and a multivariate $\pm \mathbf{p}$ (or $\pm \boldsymbol{\sigma}\cdot\mathbf{p}$) to represent these possibilities, we may represent the respective amplitudes of these four states by:

$$(ikE + i\mathbf{p} + jm)$$
$$(ikE - i\mathbf{p} + jm)$$
$$(-ikE + i\mathbf{p} + jm)$$
$$(-ikE - i\mathbf{p} + jm) \quad (17)$$

each multiplied by the same phase factor. Though the signs are intrinsically arbitrary, it will be convenient to identify these four states as representing, say,

| | |
|---|---|
| $(ikE + i\mathbf{p} + jm)$ | fermion spin up |
| $(ikE - i\mathbf{p} + jm)$ | fermion spin down |
| $(-ikE + i\mathbf{p} + jm)$ | antifermion spin down |
| $(-ikE - i\mathbf{p} + jm)$ | antifermion spin up (18) |

Once we have decided on a sign convention, however, the spin state of the particle (or, more conventionally, the helicity or handedness $\boldsymbol{\sigma}.\mathbf{p}$) is determined by the ratio of the signs of $E$ and $\mathbf{p}$. So $i\mathbf{p}$ / $ikE$ has the same helicity as $(-i\mathbf{p})$ / $(-ikE)$, but the opposite helicity to $i\mathbf{p}$ / $(-ikE)$.

In (18), the lead term may be considered as defining the fermion type. The remaining terms then are equivalent to the lead term, subjected to the respective symmetry transformations, *P*, *T* and *C*, by pre- and post-multiplication by the quaternion units defining what we will describe in section 4 as the *vacuum space*:

| | | |
|---|---|---|
| Parity | P | $\boldsymbol{i}\,(ikE + i\mathbf{p} + jm)\,\boldsymbol{i} = (ikE - i\mathbf{p} + jm)$ |
| Time reversal | T | $\boldsymbol{k}\,(ikE + i\mathbf{p} + jm)\,\boldsymbol{k} = (-ikE + i\mathbf{p} + jm)$ |
| Charge conjugation | C | $-\boldsymbol{j}\,(ikE + i\mathbf{p} + jm)\,\boldsymbol{j} = (-ikE - i\mathbf{p} + jm)$ (19) |

The charge conjugation process could equally be represented by

Charge conjugation    C    $\boldsymbol{ij}\,(ikE + i\mathbf{p} + jm)\,\boldsymbol{ij} = (-ikE - i\mathbf{p} + jm)$,

showing its ultimate origin in a vector space where $\Sigma^1\Sigma^2 = i\Sigma^3$. From (19), we can see that the rules
$$CP \equiv T,\ PT \equiv C,\ \text{and}\ CT \equiv P$$
necessarily apply, as also
$$TCP \equiv CPT \equiv \text{identity}$$
as
$$\boldsymbol{k}\,(\boldsymbol{j}\,(\boldsymbol{i}\,(ikE + i\mathbf{p} + jm)\,\boldsymbol{i})\,\boldsymbol{j})\,\boldsymbol{k} = \boldsymbol{kji}\,(ikE + i\mathbf{p} + jm)\,\boldsymbol{ijk} = (ikE + i\mathbf{p} + jm).$$

From (19) also, it will be apparent that charge conjugation is effectively defined in terms of parity and time reversal, rather than being an independent operation. This reflects the fact that only space and time are active elements, the variation in space and time being the coded information that solely determines the phase factor and the entire nature of the fermion state, and the mass term (which connects with the charge conjugation transformation) being a passive element, which can even be excluded from the operator without loss of information, as will be shown below. It is relevant here that the construction of a nilpotent amplitude effectively requires the loss of a sign degree of freedom in one component, *E*, **p** or *m*, and that the passivity of mass makes it the term to which this will apply.

There is one further refinement which allows us to reduce the amount of information needed to specify the quantum mechanics to just a two-term operator, eliminating both the mass and the phase factor. This requires a discrete or anticommutative differentiation process, with a correspondingly discrete wavefunction. Here, we define a discrete differentiation of the function *F*, which preserves the Leibniz chain rule, by taking:

$$\frac{\partial F}{\partial t} = [F, \mathcal{H}] = [F, E] \quad \text{and} \quad \frac{\partial F}{\partial X_i} = [F, P_i], \tag{20}$$

with $\mathcal{H} = E$ and $P_i$ representing energy and momentum operators [15], with the further assumption that, with velocity operators not in evidence, we may use $\partial F / \partial t$ rather than $dF / dt$. The mass term (which has only a passive role in quantum mechanics) disappears in the operator, though it is retained in the amplitude. Suppose we define a nilpotent amplitude

$$\psi = i\mathbf{k}E + i\mathbf{i}P_1 + i\mathbf{j}P_2 + i\mathbf{k}P_3 + jm$$

and an operator

$$\mathcal{D} = i\mathbf{k}\frac{\partial}{\partial t} + i\mathbf{i}\frac{\partial}{\partial X_1} + i\mathbf{j}\frac{\partial}{\partial X_2} + i\mathbf{k}\frac{\partial}{\partial X_3},$$

with

$$\frac{\partial \psi}{\partial t} = [\psi, \mathcal{H}] = [\psi, E] \quad \text{and} \quad \frac{\partial \psi}{\partial X_i} = [\psi, P_i], \tag{21}$$

After some basic algebraic manipulation, we obtain

$$-\mathcal{D}\psi = i\psi(i\mathbf{k}E + i\mathbf{i}P_1 + i\mathbf{j}P_2 + i\mathbf{k}P_3 + jm) + i(i\mathbf{k}E + i\mathbf{i}P_1 + i\mathbf{j}P_2 + i\mathbf{k}P_3 + jm)\psi - 2i(E^2 - P_1^2 - P_2^2 - P_3^2 - m^2).$$

When $\psi$ is nilpotent, then

$$\mathcal{D}\psi = \left(i\mathbf{k}\frac{\partial}{\partial t} + i\nabla\right)\psi = 0.$$

Generalising to four states, with $\mathcal{D}$ and $\psi$ represented as 4-spinors, then

$$\mathcal{D}\psi = \left(\pm i\mathbf{k}\frac{\partial}{\partial t} \pm i\nabla\right)\left(\pm i\mathbf{k}E \pm i\mathbf{i}P_1 \pm i\mathbf{j}P_2 \pm i\mathbf{k}P_3 + jm\right) = 0 \tag{22}$$

becomes the equivalent of the nilpotent Dirac equation in this discrete calculus. Significantly, the derivation of (22) does not require the $\pm i$ (or $i\hbar$) term usually applied to the differential operators in canonical quantization, though, because (22) incorporates all four sign variations, this could have been included. This not only allows a smooth transition between classical and quantum conditions but also points to an important connection between mass and the appearance of ± *i* terms in quantum mechanical operators and amplitudes.

**4 Vacuum**

Making nilpotency a universal determining factor of the fermionic state also provides us with an opportunity for understanding the concept of vacuum, with the immediate possibility of transforming from quantum mechanics to quantum field theory, without any formal process of second quantization. We simply imagine creating a fermion in some particular state (determined by added potentials, interaction terms, etc) *ab initio*, that is, from absolutely nothing. Vacuum can then be defined as the state that is left in what had previously been a complete void – that is, everything other than the fermion. So, if we define the wavefunction of the fermion as, say, $\psi_f$, the wavefunction of vacuum will be defined as $\psi_v = -\psi_f$. The superposition of fermion and vacuum will be the zero state we started from, $\psi_f + \psi_v = \psi_f - \psi_f = 0$, and, because the fermion is a nilpotent, the combination state

$$\psi_f \psi_v = -\psi_f \psi_f = -(\pm ikE \pm i\mathbf{p} + jm)(\pm ikE \pm i\mathbf{p} + jm)$$

will also be zero. In this representation, vacuum becomes the 'hole' in the zero state produced by the creation of the fermion, or, from another point of view, the 'rest of the universe' that the fermion sees and interacts with. So, if we define a fermion with interacting field terms, then the 'rest of the universe' has to be 'constructed' simultaneously to make the existence of a fermion in that state possible.

Vacuum defined in this way suggests that the universe is a zero totality, which, at the creation of every new fermion state, divides into two parts – the local fermion state and the nonlocal vacuum – and that these are connected with the simultaneous existence of two vector spaces to create the mathematical structure. The nilpotent formalism reveals that a fermion 'constructs' its own vacuum, or the entire 'universe' in which it operates, and we can consider the vacuum to be 'delocalised' to the extent that the fermion is 'localised'. If Pauli exclusion holds, no two fermions can have the same vacuum. The 'local' is now defined as whatever happens inside the nilpotent structure $(\pm ikE \pm i\mathbf{p} + jm)$, and the 'nonlocal' as whatever happens outside it. A single fermion cannot be considered as isolated but must be interacting, and construct a 'space', so that its vacuum is not localised on itself. A point-like fermion requires a dispersed vacuum. A single (noninteracting) fermion cannot exist – it can only be defined if we also define its vacuum.

Significantly, we can also show that nilpotent wavefunctions or amplitudes are also Pauli exclusive in the conventional sense of being automatically antisymmetric, with nonzero

$$\psi_1 \psi_2 - \psi_2 \psi_1 = -(\psi_2 \psi_1 - \psi_1 \psi_2)$$

since

$$(\pm ikE_1 \pm i\mathbf{p}_1 + jm_1)(\pm ikE_2 \pm i\mathbf{p}_2 + jm_2)$$
$$- (\pm ikE_2 \pm i\mathbf{p}_2 + jm_2)(\pm ikE_1 \pm i\mathbf{p}_1 + jm_1)$$
$$= 4\mathbf{p}_1\mathbf{p}_2 - 4\mathbf{p}_2\mathbf{p}_1 = 8i\,\mathbf{p}_1 \times \mathbf{p}_2. \tag{23}$$

This result implies that, instantaneously, a nilpotent wavefunction must have a **p** vector in spin space at a different orientation to any other. The instantaneous nonlocal correlation of the wavefunctions of all nilpotent could then be interpreted as the intersection of the planes corresponding to all the different **p** vector directions, and we can consider these intersections as actually creating the *meaning* of Euclidean space, with an intrinsic spherical symmetry generated by the fermions themselves.

In this context, nilpotency might be considered as simultaneously represented in two dual vector spaces, one that is rotationally symmetric, defined by $\sigma^1, \sigma^2, \sigma^3$, or the vector units **i**, **j**, **k**, and one that is rotationally asymmetric, defined by $\Sigma^1, \Sigma^2, \Sigma^3$, or the quaternion units, *i*, *j*, *k*. We can define these as the real space (the space of measurement) and the vacuum space (the space of interaction), and we will find that the nilpotent condition requires these to be dual in terms of all physical information, although presenting it in quite different forms. In the case of the vacuum space we might imagine an alternative representation of nilpotency as creating a unique direction on a set of axes defined by the values of $E$, **p** and $m$. In such a representation, half of the possibilities on one axis (those with $–m$) would be eliminated automatically (as being in the same direction as those with $m$), as would all those with zero $m$ (since the directions would all be along the line $E = p$); such hypothetical massless particles would be impossible, in addition, for fermions and antifermions with the same helicity, as $E$, $p$ has the same direction as $–E, –p$.

## 5 Spin and helicity

From the nilpotent operator ($ikE + i\mathbf{p} + jm$) we may define a Hamiltonian specified as $\mathcal{H} = –i\mathbf{k}(i\mathbf{p} + jm) = –i\mathbf{j}\mathbf{p} + iim$. If we *mathematically* define a quantity $\sigma = –\mathbf{1}$ (the pseudovector of magnitude $–1$ referred to in section 2), then

$$[\sigma, \mathcal{H}] = [–\mathbf{1}, –i\mathbf{j} (\mathbf{i}p_1 + \mathbf{j}p_2 + \mathbf{k}p_3) + iim] = [–\mathbf{1}, –i\mathbf{j}(\mathbf{i}p_1 + \mathbf{j}p_2 + \mathbf{k}p_3)]$$
$$= [–\mathbf{1}, –i\mathbf{j} (\mathbf{i}p_1 + \mathbf{j}p_2 + \mathbf{k}p_3) + iim] = [–\mathbf{1}, –i\mathbf{j}(\mathbf{i}p_1 + \mathbf{j}p_2 + \mathbf{k}p_3)]$$
$$= 2i\mathbf{j} (\mathbf{ij}p_2 + \mathbf{ik}p_3 + \mathbf{ji}p_1 + \mathbf{j}\, p_3 + \mathbf{ki}p_1 + \mathbf{kj}p_2)$$
$$= –2\mathbf{j} (\mathbf{k}(p_2 – p_1) + \mathbf{j}(p_1 – p_3) + \mathbf{i}(p_3 – p_2))$$
$$= –2j\mathbf{1} \times \mathbf{p}.$$

If **L** is an orbital angular momentum $\mathbf{r} \times \mathbf{p}$, then

$$[\mathbf{L}, \mathcal{H}] = [\mathbf{r} \times \mathbf{p}, –i\mathbf{j} (\mathbf{i}p_1 + \mathbf{j}p_2 + \mathbf{k}p_3) + iim]$$
$$= [\mathbf{r} \times \mathbf{p}, –i\mathbf{j} (\mathbf{i}p_1 + \mathbf{j}p_2 + \mathbf{k}p_3)]$$
$$= i\, [\mathbf{r}, –i\mathbf{j} (\mathbf{i}p_1 + \mathbf{j}p_2 + \mathbf{k}p_3)] \times \mathbf{p}$$

But $\qquad\qquad [\mathbf{r}, –i\mathbf{j} (\mathbf{i}p_1 + \mathbf{j}p_2 + \mathbf{k}p_3)] = i\mathbf{1}$ .

Hence $\qquad\qquad [\mathbf{L}, \mathcal{H}] = j\mathbf{1} \times \mathbf{p},$

and $\mathbf{L} + \sigma / 2$ is a constant of the motion, because

$$[\mathbf{L} + \boldsymbol{\sigma} / 2, \mathcal{H}] = 0.$$

The spin ½ term characteristic of fermionic states then emerges from this formalism purely from the multivariate properties of the **p** operator, through the additional cross product term with its imaginary coefficient or pseudovector, as in (5). Physically, it is equivalent to an intrinsic angular momentum term requiring a fermion to undergo a $4\pi$, rather than $2\pi$, rotation to return to its starting point. We can regard it as an expression of the fact that a localised point-like fermion can only be created simultaneously with a mirror image nonlocalised vacuum state. The fermion on its own gives us only half of the knowledge we require to specify the system, and this is equivalent to specifying only one of the vacuum spaces. The spin of fermion plus vacuum is, of course, single-valued (0).

We can also define helicity (**σ.p**) as another constant of the motion because

$$[\boldsymbol{\sigma}.\mathbf{p}, \mathcal{H}] = [-p, -i\mathbf{j}\,(\mathbf{i}p_1 + \mathbf{j}p_2 + \mathbf{k}p_3) + iim] = 0$$

Because, as previously specified, for a multivariate **p**,

$$\mathbf{pp} = (\boldsymbol{\sigma}.\mathbf{p})(\boldsymbol{\sigma}.\mathbf{p}) = pp = p^2,$$

we can also use **σ.p** (σp) for **p** (or **σ.∇** (σ∇) for ∇) in the nilpotent operator. A hypothetical fermion / antifermion with zero mass would be reduced to two distinguishable states:

$$\begin{aligned}(i k E + \mathbf{i}\,\boldsymbol{\sigma}.\mathbf{p} + \mathbf{j} m) &\to (i k E - i p) \\ (-i k E + \mathbf{i}\,\boldsymbol{\sigma}.\mathbf{p} + \mathbf{j} m) &\to (-i k E - i p)\end{aligned} \quad (24)$$

each of which is associated with a single sign of helicity; ($ikE + ip$) and ($-ikE + ip$) are excluded, if we choose the same sign conventions for **p**. The use of $\boldsymbol{\sigma} = -\mathbf{1}$ in deriving spin for states with positive energy ensures that the allowed spin direction for these states must be antiparallel, corresponding to left-handed helicity, with right-handed helicity for the negative energy states. Numerically, $|\pm E| = p$, so we can express the allowed states as $\pm E(\mathbf{k} - i\mathbf{i})$. Multiplication from the left by the projection operator $(1 - i\mathbf{j})/2 \equiv (1 - \gamma^5)/2$ then leaves the allowed states unchanged while zeroing the excluded ones.

Because spin has emerged in this formalism from the specifically multivariate aspect of the operator **p**, it is necessary to distinguish equations where the space variables are multivariate from those where they are not, as, for example, when polar coordinates are used. In such cases, an intrinsic spin is no longer structured into the formalism and an *explicit* spin (or total angular momentum) term has to be introduced. Dirac, however, has given a prescription for translating his equation into polar form [16], where the momentum operator acquires an additional (imaginary) spin (or total angular momentum) term, and we can easily adapt this to represent a polar transformation of the multivariate vector operator:

$$\nabla \to \left(\frac{\partial}{\partial r} + \frac{1}{r}\right) \pm i\frac{j + \tfrac{1}{2}}{r}. \quad (25)$$

and use this to define a non-time varying nilpotent operator in polar coordinates:

$$\left(ik E - ii\nabla + jm\right) \rightarrow \left(ik E - ii\left(\frac{\partial}{\partial r} + \frac{1}{r} \pm i\frac{j+\frac{1}{2}}{r}\right) + jm\right). \tag{26}$$

**6 *Zitterbewegung* and Berry phase**

A very significant aspect of spin emerges when we write a nilpotent Hamiltonian in the form

$$\mathcal{H} = -ijc\boldsymbol{\sigma}.\mathbf{p} - iiimc^2 = -ijc\mathbf{1}\mathbf{p} - iiimc^2 = \alpha c\mathbf{p} - iiimc^2,$$

with the constant *c* now specifically included (as $\hbar$ will be later), and $\alpha$ representing the original Dirac operator $\gamma\gamma^0$. Since we have four separate spin states in the system, $\alpha = -ij\mathbf{1}$ may be taken as a dynamical variable, and $\alpha c = -ij\mathbf{1}c$ defined, in terms of the discrete calculus of equations (21), as a velocity operator, which, for a free particle, becomes:

$$\mathbf{v} = \dot{\mathbf{r}} = \frac{d\mathbf{r}}{dt} = \frac{1}{i\hbar}\,[\mathbf{r},\mathcal{H}] = -ij\mathbf{1}c = c\alpha\,.$$

The use of an explicit velocity operator now requires *dF / dt* to be distinguished from $\partial F / \partial t$. The equation of motion for the velocity operator then becomes:

$$\frac{d\alpha}{dt} = \frac{1}{i\hbar}\,[\alpha,\mathcal{H}] = \frac{2}{i\hbar}(c\mathbf{p} - \mathcal{H}\alpha).$$

The solution of this well-known result, giving the equation of motion for the fermion, was first obtained by Schrödinger [17]:

$$\mathbf{r}(t) = \mathbf{r}(0) + \frac{c^2\mathbf{p}}{\mathcal{H}}t + \frac{\hbar c}{2i\,\mathcal{H}}\,[\alpha(0) - c\mathcal{H}^{-1}\mathbf{p}](exp\,(2i\mathcal{H}t\,/\,h) - 1).$$

The third term, which uniquely has no classical analogue, seemingly predicts a violent oscillatory motion or high-frequency vibration (*zitterbewegung*) of the particle at frequency $\approx 2mc^2/\hbar$, and amplitude $\hbar/2mc$, which is related to the Compton wavelength for the particle and directly determined by the particle's rest mass. Derived from a velocity operator, defined as $c\alpha = -ij\mathbf{1}c$, the *zitterbewegung* has always been interpreted as a switching between the fermion's four spin states. It is undoubtedly a vacuum effect, a continual redefinition of the localised fermion in relation to the nonlocal vacuum, without which it could not be defined at a point, and an expression of the necessity of dual vector spaces in the description of a discrete particle. It is, in effect, a direct expression of the duality of these spaces. *Zitterbewegung* describes the switching between them, a kind of gauge invariance based on their acting as carriers of the same information.

Dirac has interpreted *zitterbewegung* as implying that a fermion (or any massive particle) actually propagates along the light cone, oscillating between +*c* and –*c* at a frequency which determines its measured mass and momentum [16]. This is because a measured value of velocity can only be found by knowing positions at two different times. To find the instantaneous velocity, the time interval must be reduced to zero, thus fixing the positions with exact precision, and hence making the momentum value completely indeterminate. The ultimate significance of *zitterbewegung* in this context may be that it locates rest mass as the result of defining a singularity.

*Zitterbewegung* can thus be seen as an intrinsic aspect of defining a fermion as a point-singularity through the nilpotent structure created by dual vector spaces. A related effect can be seen in those Berry phase phenomena [18] which involve a fermion with half-integral spin subjected to a cyclic adiabatic process becoming single valued in the presence of either another fermionic state, for example, an electron (Cooper pairing) or nucleus (Jahn-Teller effect), or an 'environment' whose origin is ultimately fermionic. This could be, for example a vector potential (Aharonov-Bohm effect) or a flux line (quantum Hall effect). In each of these cases the Berry phase can be interpreted topologically, with the initial fermion travelling in a space that has changed from being simply- to multiply-connected by incorporating the other fermionic state or environment as a 'singularity'.

In fact, we could regard the unpaired fermion, defined as a pure physical singularity, as existing in its *own* multiply-connected space and thus naturally becoming a spin ½ particle. If we take bosons as products of fermion interactions, then fermions are the only known fundamental structures in nature, and experimental evidence to date suggests that they are point-like, and in this sense singularities; they are, therefore, in principle (excluding anything produced by gravity), the only known physical singularities. Now, a physical singularity can only be defined with reference to a nonlocalised phase. It is the nonlocalised phase which enables two such singularities to interact, and which allows us to describe such interactions in terms of a quantum field. In effect, information from the dual spaces of one system (potentials or even distortions of its space-time structure) creates changes in the dual spaces of the other, via changes in the $E$ and **p** terms of its operator, and, through the phase factor, of its amplitude. Even a pure vector potential (as in the Aharonov-Bohm effect) will alter the **p** term and so produce these changes. Under cyclic adiabatic conditions, we can consider the $E$ and $p$ magnitudes of the combination to be equalised as in the formation of a bosonic-type state.

## 7 Quantum mechanics and the quantum field

The nilpotent operator can be used to do ordinary relativistic quantum mechanics, if we define a probability density for a nilpotent wavefunction ($\pm i k E \pm i\mathbf{p} + jm$) by multiplication with its *complex quaternion conjugate* ($\pm i k E \mp i\mathbf{p} - jm$) (the extra 'quaternion' resulting from the fact that the nilpotent wavefunction differs from a conventional one through premultiplication by a quaternion operator). The unit probability density is then be defined by

$$\frac{(\pm i\boldsymbol{k}E \pm i\mathbf{p} + \boldsymbol{j}m)}{\sqrt{2E}} \frac{(\pm i\boldsymbol{k}E \mp i\mathbf{p} - \boldsymbol{j}m)}{\sqrt{2E}} = 1,$$

the $1/\sqrt{2E}$ being a normalizing factor. If such factors are automatically assumed to apply in calculations, we can also define $(\pm i\boldsymbol{k}E \mp i\mathbf{p} - \boldsymbol{j}m)$ as the 'reciprocal' of $(\pm i\boldsymbol{k}E \pm i\mathbf{p} + \boldsymbol{j}m)$.

Even more significantly, the nilpotent formalism intrinsically implies a full quantum field theory in which the operators act on the entire quantum field, without requiring any formal process of second quantization. The transition to quantum field theory could thus be said to occur at the point at which we choose to privilege the operator rather than the equation, and then apply Pauli exclusion to all fermionic states, whether free or bound, regardless of the number of interactions to which they are subject. A nilpotent operator, defined in this way from *absolutely nothing*, then becomes a creation operator acting on vacuum to create the fermion, together with all the interactions in which it is involved. No further mathematical formalism is necessary, and neither quantum mechanics nor quantum field theory requires specification by an equation. Once the operator is defined, the phase factor then becomes an expression of all the possible variations in space and time which are encoded in the operator, and is uniquely defined with it. A fermion is thus specified as a set of space and time variations, with the mass term a purely passive quantity, and convenient, rather than necessary information.

Calculations are notably easier and more efficient than those of alternative formalisms, largely because *dual* information, concerning both fermion and vacuum, is available, as is completely new *physical* information. For example, if we take the use of polar coordinates as representing spherical symmetry with respect to a point source, then the operator (26) has no nilpotent solutions unless the *E* term also contains an expression proportional to $1/r$. It would seem that simply defining a point source forces us to assume that a Coulomb interaction component is necessary for any nilpotent fermion defined with respect to it. All known forces have such components, together with an associated $U(1)$ symmetry. For the gravitational and electric forces, it is the main or complete description; for the strong force it is the one-gluon exchange; for the weak field it is the hypercharge and the $B^0$ gauge field. Its effect is connected purely with scale or magnitude and we can associate it with the coupling constant.

If we now write the nilpotent operator in (26) with the required Coulomb term, we will find that it can be solved, using the known procedures, but eliminating many unnecessary ones, in only six lines of calculation. We begin with:

$$\left( \pm i\boldsymbol{k}\left(E - \frac{A}{r}\right) \mp i\boldsymbol{i}\left(\frac{\partial}{\partial r} + \frac{1}{r} \pm i\frac{j+\tfrac{1}{2}}{r}\right) + \boldsymbol{j}m \right). \qquad (27)$$

with a main requirement to find the phase factor $\phi$ which will make the amplitude nilpotent. So, we try the standard solution:

$$\phi = e^{-ar} r^\gamma \sum_{\nu=0} a_\nu r^\nu .$$

We then apply the operator in (24) to $\phi$, and square the result to 0 to obtain:

$$4\left(E-\frac{A}{r}\right)^2 = -2\left(-a+\frac{\gamma}{r}+\frac{v}{r}+...\frac{1}{r}+i\frac{j+\frac{1}{2}}{r}\right)^2 -2\left(-a+\frac{\gamma}{r}+\frac{v}{r}+...\frac{1}{r}-i\frac{j+\frac{1}{2}}{r}\right)^2 +4m^2.$$

Equating constant terms leads to

$$a = \sqrt{m^2 - E^2}. \qquad (28)$$

Equating terms in $1/r^2$, following standard procedure, with $v = 0$, we obtain:

$$\left(\frac{A}{r}\right)^2 = -\left(\frac{\gamma+1}{r}\right)^2 + \left(\frac{j+\frac{1}{2}}{r}\right)^2. \qquad (29)$$

Assuming the power series terminates at $n'$, following another standard procedure, and equating coefficients of $1/r$ for $v = n'$,

$$2EA = -2\sqrt{m^2 - E^2}\,(\gamma + 1 + n'), \qquad (30)$$

the terms in $(j + \frac{1}{2})$ cancelling over the summation of the four multiplications, with two positive and two negative. Algebraic rearrangement of (28)-(30) then yields

$$\frac{E}{m} = \frac{1}{\sqrt{1+\dfrac{A^2}{(\gamma+1+n')^2}}} = \frac{1}{\sqrt{1+\dfrac{A^2}{\left(\sqrt{(j+\frac{1}{2})^2 - A^2}+n'\right)^2}}},$$

With $A = Ze^2$, this becomes the hyperfine or fine structure formula for a one-electron nuclear atom or ion (e.g. the hydrogen atom, where $Z = 1$).

**8 Bosons**

From (19), we have seen that the terms in the nilpotent 4-spinor, other than the lead term – which determines the nature of the 'real' particle state in *real* space or the space of observation – are effectively, the *P*-, *T*- and *C*-transformed versions of this state, the states into which it could transform without changing the magnitude of its energy or momentum. We can also perceive them as vacuum 'reflections' of the real particle state, and we will show in section 10 how they arise from vacuum operations that can be mathematically defined, through a partitioning of the continuous vacuum into a 3-dimensional vacuum space, with each reflection being in one 'dimension' of the space. Now, although Pauli exclusion prevents a fermion from forming a combination state with itself, we can imagine it forming a combination state with each of these vacuum 'reflections', and, if the 'reflection' exists or materialises as a 'real' state, then the combined state can form one of the three classes of bosons or boson-like objects.

A combination of fermion and antifermion with the same spins but opposite helicities produces a state equivalent to a spin 1 boson. We take, for example, the product of a row vector fermion and a column vector antifermion, both written as columns for convenience:

$$\begin{aligned}(ikE + i\,\mathbf{p} + j\,m) &\quad (-ikE + i\,\mathbf{p} + j\,m) \\ (ikE - i\,\mathbf{p} + j\,m) &\quad (-ikE - i\,\mathbf{p} + j\,m) \\ (-ikE + i\,\mathbf{p} + j\,m) &\quad (ikE + i\,\mathbf{p} + j\,m) \\ (-ikE - i\,\mathbf{p} + j\,m) &\quad (ikE - i\,\mathbf{p} + j\,m).\end{aligned} \qquad (31)$$

While the antifermion structure reverses the signs of $E$ throughout, and spin reversal changes the sign of $\mathbf{p}$, the phase factor of both fermion and antifermion components is, according to our original construction of the nilpotent formalism, the same, dependent on the values of $E$ and $\mathbf{p}$ but not on their signs. The sign variations ensure cancellation of all the terms with quaternion coefficients, so the product is a nonzero scalar, and the same result will be obtained if the spin 1 boson is massless (as is the case with such gauge bosons as photons and gluons). Then we have:

$$\begin{aligned}(ikE + i\,\mathbf{p}) &\quad (-ikE + i\,\mathbf{p}) \\ (ikE - i\,\mathbf{p}) &\quad (-ikE - i\,\mathbf{p}) \\ (-ikE + i\,\mathbf{p}) &\quad (ikE + i\,\mathbf{p}) \\ (-ikE - i\,\mathbf{p}) &\quad (ikE - i\,\mathbf{p}).\end{aligned} \qquad (32)$$

The spin 0 boson structure is obtained by reversing the $\mathbf{p}$ signs in either fermion or antifermion, so that the components have the opposite spins but the same helicities:

$$\begin{aligned}(ikE + i\,\mathbf{p} + j\,m) &\quad (-ikE - i\,\mathbf{p} + j\,m) \\ (ikE - i\,\mathbf{p} + j\,m) &\quad (-ikE + i\,\mathbf{p} + j\,m) \\ (-ikE + i\,\mathbf{p} + j\,m) &\quad (ikE - i\,\mathbf{p} + j\,m) \\ (-ikE - i\,\mathbf{p} + j\,m) &\quad (ikE + i\,\mathbf{p} + j\,m).\end{aligned} \qquad (33)$$

Again the product is a nonzero scalar, but, in this case, reducing the mass will zero the product as well.

$$\begin{aligned}(ikE + i\,\mathbf{p}) &\quad (-ikE - i\,\mathbf{p}) = 0 \\ (ikE - i\,\mathbf{p}) &\quad (-ikE + i\,\mathbf{p}) = 0 \\ (-ikE + i\,\mathbf{p}) &\quad (ikE - i\,\mathbf{p}) = 0 \\ (-ikE - i\,\mathbf{p}) &\quad (ikE + i\,\mathbf{p}) = 0\end{aligned} \qquad (34)$$

The implication is that a spin 0 boson, defined by this process, cannot be massless. So Goldstone bosons cannot exist, and the Higgs boson must have a mass. The mass is, additionally, as will become apparent, a measure of the degree of right-handedness in the fermion component and left-handedness in the antifermion component.

A third type of boson-like state can be formed by combining two fermions with opposite spins and opposite helicities:

$$(ikE + i\mathbf{p} + jm) \quad (kE - i\mathbf{p} + jm)$$
$$(ikE - i\mathbf{p} + jm) \quad (ikE + i\mathbf{p} + jm)$$
$$(-ikE + i\mathbf{p} + jm) \quad (-ikE - i\mathbf{p} + jm)$$
$$(-ikE - i\mathbf{p} + jm) \quad (-ikE + i\mathbf{p} + jm). \quad (35)$$

States of this kind can be imagined to occur in Cooper pairing in superconductors, in $He^4$ and Bose-Einstein condensates, in spin 0 nuclei, in the Jahn-Teller effect, the Aharonov-Bohm effect, the quantum Hall effect, and, in general, in states where there is a nonzero Berry phase to make fermions become single-valued in terms of spin. In most cases, these will be spin 0 states, but spin 1 fermion-fermion combinations will be possible if, as is the case with $He^3$, the two components move with respect to each other with components of motion in opposite directions, presumably in some kind of harmonic oscillator fashion, meaning that they could have the same spin states but opposite helicities. If they are spin 0, they can also have zero effective mass, as in Cooper pairing.

Now, the weak interaction can be considered as one in which fermions and antifermions are annihilated while bosons are created, or bosons are annihilated while fermions and antifermions are created, and, more generally, as one in which both processes (or equivalent) occur. As a creator and annihilator of states, it has the action of a harmonic oscillator. One of the fundamental differences between fermions and bosons is that fermions are sources for weak interactions, while bosons are not. Bosons, considered as created at fermion-antifermion vertices, are the products of weak interactions. Even in examples such as electron-positron collisions, where the predominant interaction is electric at low energies, there is an amplitude for a weak interaction. If we consider (32)-(35) as defining the vertices for boson production via the weak interaction, then it appears from (32) and from (24) that the pure weak interaction requires left-handed fermions and right-handed antifermions. In other words, it requires both a charge-conjugation violation and a simultaneous parity or time-reversal violation.

We can see in principle how this leads to mass generation by some process at least resembling the Higgs mechanism. Suppose we imagine a fermionic vacuum state with zero mass, say ($ikE + i\mathbf{p}$). An ideal vacuum would maintain exact and absolute *C*, *P* and *T* symmetries. Under *C* transformation, ($ikE + i\mathbf{p}$) would become ($-ikE - i\mathbf{p}$), with which it would be indistinguishable under normalization. No bosonic state would be required for the transformation, because the states would be identical. If, however, the vacuum state is degenerate in some way under charge conjugation (as supposed in the weak interaction), then ($ikE + i\mathbf{p}$) will be transformable into a state which can be distinguished from it, and the bosonic state ($ikE + i\mathbf{p}$) ($-ikE - i\mathbf{p}$) will necessarily exist. However, this can only be true if the state has nonzero mass and becomes the spin 0 'Higgs boson' ($ikE + i\mathbf{p} + jm$) ($-ikE - i\mathbf{p} + jm$). The mechanism, which produces this state, and removes the masslessness of the boson, requires the fixing of a gauge for the weak interaction (a 'filled' weak vacuum), which manifests itself in the massive intermediate bosons, *W* and *Z*.

The structures of bosons and the consideration of spin in section 5 suggest that mass and helicity are closely related. If the degree of left-handed helicity is determined by the ratio (±) *i***p** / (±) *ikE*, then the addition of a mass term will change this ratio. Similarly, a change in the helicity ratio will also affect the mass. If the weak interaction is only responsive to left-handed helicity states in fermions, then right-handed states will be intrinsically passive, so having no other function except to generate mass. The presence of two helicity states will be a signature of the presence of mass. The *SU*(2) of weak isospin, which, in effect, expresses the invariability of the weak interaction to the addition of an opposite degree of helicity (due to the presence of, say, mass or electric charge) is thus related indirectly to the *SU*(2) of spin, which is a simple description of the existence of two helicity states. It is significant that the *zitterbewegung* frequency, which is a measure of the switching of helicity states, depends only on the fermion's mass. Mass is in some sense created by it, or is in some sense an expression of it. The restructuring of space and time variation or energy and momentum, via the phase factor, during an interaction, leads to a creation or annihilation of mass, which manifests itself in the restructuring of the *zitterbewegung*.

The coupling of a massless fermion, say (*ikE*$_1$ + *i***p**$_1$), to a Higgs boson, say (*ikE* + *i***p** + *jm*) (−*ikE* − *i***p** + *jm*), to produce a massive fermion, say (*ikE*$_2$ + *i***p**$_2$ + *jm*$_2$), can be imagined as occurring at a vertex between the created fermion (*ikE*$_2$ + *i***p**$_2$ + *jm*$_2$) and the antistate (−*ikE*$_1$ − *i***p**$_1$), to the annihilated massless fermion, with subsequent equalization of energy and momentum states. If we imagine a vertex involving a fermion superposing (*ikE* + *i***p** + *jm*) and (*ikE* − *i***p** + *jm*) with an antifermion superposing (−*ikE* + *i***p** + *jm*) and (−*ikE* − *i***p** + *jm*), then there will be a minimum of two spin 1 combinations and two spin 0 combinations, meaning that the vertex will be massive (with Higgs coupling) and carry a non-weak (i.e. electric) charge. So, a process such as a weak isospin transition, which, to use a very basic model, converts something like (*ikE*$_1$ + *i***p**$_1$ + *jm*$_1$) (representing isospin up) to something like $\alpha_1$ (*ikE*$_2$ + *i***p**$_2$ + *jm*$_2$) + $\alpha_1$ (*ikE*$_2$ − *i***p**$_2$ + *jm*$_2$) (representing isospin down), requires an additional Higgs boson vertex (spin 0) to accommodate the right-handed part of the isospin down state, when the left-handed part interacts weakly. This is, of course, what we mean when we say that the *W* and *Z* bosons have mass. The mass balance is done through separate vertices involving the Higgs boson.

One further consideration leads to a prediction in the case of the fermion-fermion spin 0 state. Because the formation of the spin 0 state necessarily requires intrinsically massive components, even in those cases where it assumes nonzero effective mass through a Fermi velocity less than *c*, time reversal symmetry (the one applicable to the transition) must be broken in the weak formation or decay of such states. The most likely opportunity of observing such a process might be in one of the physical manifestations of the nonzero Berry phase, say the quantum Hall effect, in some special type of condensed matter such as graphene. Here, the conduction electrons have zero effective mass and a Hamiltonian that can be written in the form ± $v_F$***i***(**i**$p_x$ + **j**$p_y$), where $v_F$ is the Fermi velocity. We can imagine creating a boson-like state with single-valued spin by the quantum Hall effect, Aharonov-Bohm effect, or Bose-Einstein condensation, and then observing, perhaps through a change in the Fermi velocity during its decay, the violation of both *P* and *CP* = *T* symmetries.

## 9 Baryons

Fermions are defined as singularities through their dual space structures. Baryons complicate this structure by introducing an explicit 3-dimensionality into the real space part of the structure. They are an expression of the fact that singularities are incompatible with a pure 3-dimensional space and are only possible where we have a dual space. Though we clearly cannot combine three components in the form:

$$(ikE \pm i\,\mathbf{p} + j\,m)\,(ikE \pm i\,\mathbf{p} + j\,m)\,(ikE \pm i\,\mathbf{p} + j\,m)$$

as this will automatically reduce to zero, we can imagine a three-component structure in which the vector nature of $\mathbf{p}$ plays an explicit role

$$(ikE \pm i\,\mathbf{i}p_x + j\,m)\,(ikE \pm i\,\mathbf{j}p_y + j\,m)\,(ikE \pm i\,\mathbf{k}p_z + j\,m)$$

and which has nilpotent solutions when $\mathbf{p} = \pm i\,\mathbf{i}p_x$, $\mathbf{p} = \pm i\,\mathbf{j}p_y$, or $\mathbf{p} = \pm i\,\mathbf{k}p_z$, or when the momentum is directed entirely along the *x*, *y*, or *z* axes, in either direction, however arbitrarily these are defined. The complete wavefunction will, in effect, contain information from the equivalent of six allowed independent nonlocally gauge invariant phases, all existing simultaneously and subject to continual transitions at a constant rate:

$$
\begin{array}{lll}
(ikE + i\,\mathbf{i}p_x + j\,m)\;(ikE + \ldots + j\,m)\;(ikE + \ldots + j\,m) & +RGB \\
(ikE - i\,\mathbf{i}p_x + j\,m)\;(ikE - \ldots + j\,m)\;(ikE - \ldots + j\,m) & -RBG \\
(ikE + \ldots + j\,m)\;(ikE + i\,\mathbf{j}p_y + j\,m)\;(ikE + \ldots + j\,m) & +BRG \\
(ikE - \ldots + j\,m)\;(ikE - i\,\mathbf{j}p_y + j\,m)\;(ikE - \ldots + j\,m) & -GRB \\
(ikE + \ldots + j\,m)\;(ikE + \ldots + j\,m)\;(ikE + i\,\mathbf{k}p_z + j\,m) & +GBR \\
(ikE - \ldots + j\,m)\;(ikE - \ldots + j\,m)\;(ikE - i\,\mathbf{k}p_z + j\,m) & -BGR \quad (36)
\end{array}
$$

Using an appropriate normalization, these reduce to

$$
\begin{array}{ll}
(ikE + i\,\mathbf{i}p_x + j\,m) & +RGB \\
(ikE - i\,\mathbf{i}p_x + j\,m) & -RBG \\
(ikE - i\,\mathbf{j}p_y + j\,m) & +BRG \\
(ikE + i\,\mathbf{j}p_y + j\,m) & -GRB \\
(ikE + i\,\mathbf{k}p_z + j\,m) & +GBR \\
(ikE - i\,\mathbf{k}p_z + j\,m) & -BGR \quad (37)
\end{array}
$$

with the third and fourth changing the sign of the $\mathbf{p}$ component. The group structure required to maintain these phases is an *SU*(3) structure, with eight generators and wavefunction, exactly as in the conventional model using coloured quarks,

$$\psi \sim (BGR - BRG + GRB - GBR + RBG - RGB).$$

'Colour' transitions in (36) are produced either by an exchange of the components of **p** between the individual quarks or baryon components, or by a relative switching of the component positions. No direction is privileged, so the transition must be gauge invariant, and the mediators must be massless, exactly as with the eight massless gluons of the gluon structure. Here, six gluons are constructed from:

$$(\pm kE \mp ii\, \mathbf{i}p_x)\,(\mp kE \mp ii\, \mathbf{j}p_y) \quad (\pm kE \mp ii\, \mathbf{j}p_y)\,(\mp kE \mp ii\, \mathbf{i}p_x)$$
$$(\pm kE \mp ii\, \mathbf{j}p_y)\,(\mp kE \mp ii\, \mathbf{k}p_z) \quad (\pm kE \mp ii\, \mathbf{k}p_z)\,(\mp kE \mp ii\, \mathbf{j}p_y)$$
$$(\pm kE \mp ii\, \mathbf{i}p_z)\,(\mp kE \mp ii\, \mathbf{i}p_x) \quad (\pm kE \mp ii\, \mathbf{i}p_x)\,(\mp kE \mp ii\, \mathbf{i}p_z) \qquad (38)$$

and two from combinations of

$$(\pm kE \mp ii\, \mathbf{i}p_x)\,(\mp kE \mp ii\, \mathbf{i}p_x) \quad (\pm kE \mp ii\, \mathbf{j}p_y)\,(\mp kE \mp ii\, \mathbf{j}p_y)$$
$$(\pm kE \mp ii\, \mathbf{k}p_z)\,(\mp kE \mp ii\, \mathbf{k}p_z) \qquad (39)$$

A representation such as (36), which shows only one 'quark' active at any time in contributing to the angular momentum operator, indicates clearly why only 1/3 of baryon spin has been found to be due to the valence quarks. The rest of the spin then becomes a 'vacuum' contribution, split approximately 3 to 1 in favour of the gluons over the sea quarks, the gluons thus taking half the overall total.

The structures derived in this section produce insights into at least two fundamental physical problems. The first is the mass-gap problem for baryons. Here, we are confronted with the fact that baryons have nonzero mass and yet this mass is thought to be produced by the action of massless gluons. In addition, although the Higgs mechanism appears to be the main process by which mass is delivered to fermions, the gluon exchange is generally considered to be a non-Higgs process. In fact, the structures in (36) clearly require the simultaneous existence of two states of helicity for the symmetry to remain unbroken, and this can only be possible if the baryon has nonzero mass. Further, this process is the signature of the Higgs mechanism, and so, contrary to much current supposition, the generation of the masses of baryons follows exactly the same process as that of all other fermions. However, this does not contradict the fact, established by much calculation using QCD, that the bulk of the mass of a baryon is due to the exchange of massless gluons, as the exchange of gluons structured as in (38) and (39) will necessarily lead to a sign change in the **p** operator, and hence of helicity, the exact mechanism which is responsible for the production of all known particle masses. In fact, the same will be true of all fermions involved in spin 1 boson exchange, and so all fermions must have nonzero masses.

The second problem is the specific nature and mechanism of the strong interaction between quarks. Again, we see that a solution is suggested by the exact structure of the nilpotent operator. Here, we know, from (26), that there must be a Coulomb component or inverse linear potential ($\propto 1/r$), just to accommodate spherical symmetry. This has a known physical manifestation in the one-gluon exchange. But there is also at least one other component, which is responsible for quark confinement, for infrared slavery and for asymptotic freedom, and a linear potential ($\propto r$) has long been hypothesized and used in calculations. Here, we see that an exchange of **p** components at a constant rate, as in (36), would, in principle, require a constant rate of change of momentum, which is the signature of a linear potential.

In the nilpotent formalism, a differential operator incorporating Coulomb and linear potentials from a source with spherical symmetry (either the centre of a 3-quark system or one component of a quark-antiquark pairing) can be written in the form:

$$\left( \pm k \left( E + \frac{A}{r} + Br \right) \mp i \left( \frac{\partial}{\partial r} + \frac{1}{r} \pm i \frac{j + \frac{1}{2}}{r} \right) + ijm \right). \tag{40}$$

If we can identify the phase factor to which this operator applies, to yield nilpotent solutions, it might be possible to show, for the first time on an analytic basis, that it is associated with a force which has characteristics identifiable with those of the strong interaction. By analogy with the pure Coulomb calculation, we might propose that the phase factor is of the form:

$$\phi = \exp(-ar - br^2) r^\gamma \sum_{\nu=0} a_\nu r^\nu ,$$

Applying the operator in (38) and the nilpotent condition, we obtain:

$$E^2 + 2AB + \frac{A^2}{r^2} + B^2 r^2 + \frac{2AE}{r} + 2BEr = m^2$$

$$- \left( a^2 + \frac{(\gamma + \nu + ... + 1)^2}{r^2} - \frac{(j + \frac{1}{2})^2}{r^2} + 4b^2 r^2 + 4abr - 4b(\gamma + \nu + ... + 1) - \frac{2a}{r}(\gamma + \nu + ... + 1) \right)$$

with the positive and negative $i(j + \frac{1}{2})$ terms cancelling out over the four solutions, as previously. Then, assuming a termination in the power series (as with the Coulomb solution), we can equate:

        coefficients of $r^2$ to give          $B^2 = -4b^2$
        coefficients of $r$ to give          $2BE = -4ab$
        coefficients of $1/r$ to give          $2AE = 2a(\gamma + \nu + 1)$

These equations immediately lead to:

$$b = \pm \frac{iB}{2}$$
$$a = \mp iE$$
$$\gamma + \nu + 1 = \mp iA .$$

The ground state case (where $\nu = 0$) then requires a phase factor of the form:

$$\phi = \exp(\pm iEr \mp iBr^2 / 2) r^{\mp iqA - 1} .$$

The imaginary exponential terms in $\phi$ can be seen as representing asymptotic freedom, the exp ($\mp iEr$) being typical for a free fermion. The complex $r^\gamma$ term can be structured as a component phase, $\chi(r) = \exp(\pm iqA \ln(r))$, which varies less rapidly with $r$ than the rest of $\phi$. We can therefore write $\phi$ as

$$\phi = \frac{\exp(kr + \chi(r))}{r},$$

where
$$k = \pm iE \mp iBr/2.$$

The first term dominates at high energies, where $r$ is small, approximating to a free fermion solution, which can be interpreted as asymptotic freedom, while the second term, with its confining potential $Br$, dominates, at low energies, when $r$ is large, and this can be interpreted as infrared slavery. The Coulomb term, which is required to maintain spherical symmetry, is the component which defines the strong interaction phase, $\chi(r)$, and this can be related to the directional status of **p** in the state vector.

## 10 Partitioning the vacuum

The nilpotent formalism defines a continuous vacuum $-(\pm ikE \pm i\mathbf{p} + j m)$ to each fermion state $(\pm ikE \pm i\mathbf{p} + j m)$, and this vacuum expresses the nonlocal aspect of the state. However, the use of the operators ***k***, ***i***, ***j*** suggests that we can partition this state into discrete components with a dimensional structure. If we postmultiply $(\pm ikE \pm i\mathbf{p} + j m)$ by the idempotent ***k***$(\pm ikE \pm i\mathbf{p} + j m)$ any number of times, the only change is to introduce a scalar multiple, which can be normalized away.

$$(\pm ikE \pm i\mathbf{p} + j m)\, k(\pm ikE \pm i\mathbf{p} + j m)\, k(\pm ikE \pm i\mathbf{p} + j m) \dots \rightarrow (\pm ikE \pm i\mathbf{p} + j m) \quad (41)$$

The idempotent acts as a vacuum operator. The same applies to postmultiplication by ***i***($\pm ikE \pm i\mathbf{p} + j m$) or ***j***($\pm ikE \pm i\mathbf{p} + j m$), except that the latter also produces a unit vector which disappears on every alternate postmultiplication. However, the operation expressed in (41) is also equivalent to applying a time-reversal transformation to every even ($\pm ikE \pm i\mathbf{p} + j m$). Then we have

$$(\pm ikE \pm i\mathbf{p} + j m)\,(\mp ikE \pm i\mathbf{p} + j m)\,(\pm ikE \pm i\mathbf{p} + j m) \dots \rightarrow (\pm ikE \pm i\mathbf{p} + j m) \quad (42)$$

with every alternate state becoming an antifermion, which combines with the original fermion state to become a spin 1 boson ($\pm ikE \pm i\mathbf{p} + j m$) ($\mp kE \pm i\mathbf{p} + j m$).

If we apply the same process using ***i***($\pm ikE \pm i\mathbf{p} + j m$) and ***j***($\pm ikE \pm i\mathbf{p} + j m$), we obtain results which suggest that, from an initial fermion state, we can generate either three vacuum reflections, via respective *T*, *P* and *C* transformations, representing antifermion with the same spin, fermion with opposite spin, and antifermion with opposite spin, or combined particle-vacuum states which have the respective structures of spin 1 bosons, spin 0 bosons, or boson-like paired fermion (PF) combinations of the same kind as constitute Cooper pairs and the elements of Bose-Einstein condensates. Using just the lead terms of the nilpotents, and assuming that we can complete the spinor structures using the 3 conventional sign variations, we could represent these as:

$$(ikE + i\mathbf{p} + jm)\, k\, (ikE + i\mathbf{p} + jm)\, k\, (ikE + i\mathbf{p} + jm)\, k\, (ikE + i\mathbf{p} + jm) \ldots \quad T$$
$$(ikE + i\mathbf{p} + jm)\, (-ikE + i\mathbf{p} + jm)\, (ikE + i\mathbf{p} + jm)\, (-ikE + i\mathbf{p} + jm) \ldots \quad \text{spin 1}$$

$$(ikE + i\mathbf{p} + jm)\, j\, (ikE + i\mathbf{p} + jm)\, j\, (ikE + i\mathbf{p} + jm)\, j\, (ikE + i\mathbf{p} + jm) \ldots \quad P$$
$$(ikE + i\mathbf{p} + jm)\, (-ikE - i\mathbf{p} + jm)\, (ikE + i\mathbf{p} + jm)\, (-ikE - i\mathbf{p} + jm) \ldots \quad \text{spin 0}$$

$$(ikE + i\mathbf{p} + jm)\, i\, (ikE + i\mathbf{p} + jm)\, i\, (ikE + i\mathbf{p} + jm)\, i\, (ikE + i\mathbf{p} + jm) \ldots \quad C$$
$$(ikE + i\mathbf{p} + jm)\, (ikE - i\mathbf{p} + jm)\, (ikE + i\mathbf{p} + jm)\, (ikE - i\mathbf{p} + jm) \ldots \quad \text{PF} \qquad (43)$$

According to these processes, repeated post-multiplication of a fermion operator by any of the discrete idempotent vacuum operators creates an alternate series of antifermion and fermion vacuum states, or, equivalently, an alternate series of boson and fermion states without changing the character of the real particle state. A fermion produces a boson state by combining with its own vacuum image, and the two states form a supersymmetric partnership. Nilpotent operators are thus intrinsically supersymmetric, with supersymmetry operators typically of the form:

$$\text{Boson to fermion:} \quad Q = (\pm ikE \pm i\mathbf{p} + jm)$$
$$\text{Fermion to boson:} \quad Q^\dagger = (\mp ikE \pm i\mathbf{p} + jm)$$

A fermion converts to a boson by multiplication by an antifermionic operator $Q^\dagger$; a boson converts to a fermion by multiplication by a fermionic operator $Q$, and we can represent the first sequence in (42) by the supersymmetric

$$Q\, Q^\dagger\, Q\, Q^\dagger\, Q\, Q^\dagger\, Q\, Q^\dagger\, Q \ldots$$

It may be that we can choose to interpret this as the series of boson and fermion loops, of the same energy and momentum, required by the exact supersymmetry which would eliminate the need for renormalization, and remove the hierarchy problem altogether. Fermions and bosons (with the same values $E$, $\mathbf{p}$ and $m$) then become their own supersymmetric partners through the creation of vacuum states, making the hypothesis of a set of real supersymmetric particles to solve the hierarchy problem potentially superfluous.

The identification of $i(ikE + i\mathbf{p} + jm)$, $k(ikE + i\mathbf{p} + jm)$ and $j(ikE + i\mathbf{p} + jm)$ as vacuum operators and $(ikE - i\mathbf{p} + jm)$, $(-ikE + i\mathbf{p} + jm)$ and $(-ikE - i\mathbf{p} + jm)$ as their respective vacuum 'reflections' at interfaces provided by $P$, $T$ and $C$ transformations suggests a new insight into the meaning of the Dirac 4-spinor. With the extra knowledge we have now gained, we can interpret the three terms other than the lead term *in the spinor* as the vacuum 'reflections' that are created with the particle. We can regard the existence of three vacuum operators as a result of a partitioning of the vacuum as a result of quantization and as a consequence of the 3-part structure observed in the nilpotent fermionic state, while the *zitterbewegung* can be taken as an indication that the vacuum is active in defining the fermionic state.

Taken together, the four components of the spinor cancel exactly, especially when represented as operators using discrete calculus, as in (21). (This is, of course, for the operator or amplitude in nilpotent mode; in idempotent mode, the summed amplitudes would normalize to 1.) The four components can be represented as creation operators for

| | |
|---|---|
| fermion spin up | $(ikE + i\mathbf{p} + jm)$ |
| fermion spin down | $(ikE - i\mathbf{p} + jm)$ |
| antifermion spin down | $(-ikE + i\mathbf{p} + jm)$ |
| antifermion spin up | $(-ikE - i\mathbf{p} + jm)$ |

or annihilation operators for

| | |
|---|---|
| antifermion spin down | $(ikE + i\mathbf{p} + jm)$ |
| antifermion spin up | $(ikE - i\mathbf{p} + jm)$ |
| fermion spin up | $(-ikE + i\mathbf{p} + jm)$ |
| fermion spin down | $(-ikE - i\mathbf{p} + jm)$ |

They could equally well be regarded as two operators for creation and two for annihilation, for example:

| | |
|---|---|
| fermion spin up creation | $(ikE + i\mathbf{p} + jm)$ |
| fermion spin down creation | $(ikE - i\mathbf{p} + jm)$ |
| fermion spin up annihilation | $(-ikE + i\mathbf{p} + jm)$ |
| fermion spin down annihilation | $(-ikE - i\mathbf{p} + jm)$ |

Either way, the cancellation is exact, both physically, and algebraically (when we use the discrete operators which leave out the passive mass component). It is interesting that the cancellation requires *four* components, rather than two, for, while the transitions:

$$(ikE + i\mathbf{p} + jm) \rightarrow (ikE - i\mathbf{p} + jm)$$
and
$$(ikE + i\mathbf{p} + jm) \rightarrow (-ikE + i\mathbf{p} + jm)$$

can occur through spin 1 boson and spin 0 paired fermion exchange, and the active space and time components, there is no process in nature for the *direct* transition:

$$(ikE + i\mathbf{p} + jm) \rightarrow (-ikE - i\mathbf{p} + jm)$$

with no active component as agent. In this context, it might be worth noting that the spin 0 fermion-fermion state

$$(ikE + i\mathbf{p} + jm)(ikE - i\mathbf{p} + jm)$$

is such as would be required in a pure weak transition from $-ikE$ to $+ikE$, or its inverse.

We can also see the three vacuum coefficients **k**, **i**, **j** as originating in (or being responsible for) the concept of discrete (point-like) charge. In effect, the operators, **k**, **i** and **j** perform the functions of weak, strong and electric 'charges' or sources, acting to partition the *continuous* vacuum represented by $-(i k E + i\mathbf{p} + j m)$, and responsible for zero-point energy, into discrete components, whose special characteristics are determined by the respective pseudoscalar, vector and scalar natures of their associated terms $iE$, **p** and $m$.

In this sense, they are related to 'real' weak, strong and electric localized charges, though they are delocalized. We can describe the partitions as strong, weak and electric 'vacua', and assign to them particular roles within existing physics:

| | | |
|---|---|---|
| **k** ($ikE + i\mathbf{p} + jm$) | weak vacuum | fermion creation |
| **i** ($ikE + i\mathbf{p} + jm$) | strong vacuum | gluon plasma |
| **j** ($ikE + i\mathbf{p} + jm$) | electric vacuum | isospin / hypercharge |

These three vacua retain the characteristics of the generating charge structures, respectively pseudoscalar, vector and scalar, which explain also the special characteristics and group structures of the forces with which they are associated. It is the vector characteristic of the strong vacuum that makes baryon structure possible, and it is the pseudoscalar characteristic of the weak vacuum that makes the link between particle structure and vacuum possible at all. The 'electric vacuum' – empty or filled – can be seen as responsible for the transition between weak isospin up and down states.

The total vacuum $-1(\pm ikE \pm i\mathbf{p} + jm)$, which is partitioned by the **k**, **i**, **j** operators, can be thought of as the continuous gravitational vacuum (with negative energy), which supplies the mechanism for the instantaneous transmission of quantum correlation, and which ensures that the nilpotent mechanism accommodates the operation of both quantum holography, where $E$ and $m$ become the phase and reference phase, and the holographic principle, where the $E$ and **p** terms create the effective 'bounding area' [1]. The holographic aspects are dual to the nilpotent aspects, and can be observed directly by reversing the roles of vectors (connected to space) and quaternions (connected to charge) in the nilpotent structure. The holographic information will then determine the nature of the system, including connected information about its inertial mass and charge structure. For example, just as the electric charge determines the inertia of the electron, via the holographic principle [1], so the strong charge determines the inertia of the first generation bare quarks at about 3 to 6 MeV, and the weak charge seemingly determines the inertia of the lightest neutrino at something like 0.13 eV.

The special nature of the gravitational vacuum as a kind of 'sum' of the others has been a fundamental component of nilpotent quantum mechanics from the beginning and is also present in its antecedent theories. This has a direct connection with the gravity / gauge theory correspondence which has now appeared in string theory. Essentially, gravity and gauge theory (strong and electroweak) are dual. This is also evident in the way that the holographic principle privileges gravity to obtain information about the entire system. One can be used to provide information about the other. The fundamental duality is that of the nonlocal (gravity) and local (gauge theory), which tells us why gravity is so weak and why it is not obviously a quantized force. (My own work has approached gravitational quantization through the inertial reaction. [1]) It also appears to tell us why the 'cosmological constant' is at the opposite end of the possible physical scale (in information terms) to the one worked out from quantum gravity. In principle, the duality at the heart of quantum physics says that the local is impossible without the nonlocal; real space requires its dual in vacuum space.

## 11 A perturbation calculation

If exact supersymmetry is a consequence of the nilpotent formalism and its representation of vacuum, then a free fermion in vacuum should produce its own loop cancellations and its energy should acquire a finite value without renormalization. Free fermion plus boson loops should cancel, and there should be no hierarchy problem. We can examine this possibility by performing a basic perturbation calculation for first order coupling in QED, and showing that it leads to zero in the case of a free fermion. Suppose we have a fermion acted on by the electromagnetic potentials $\phi$, $\mathbf{A}$. Then, using only the lead terms of the spinors for simplicity,

$$\left(-\mathbf{k}\frac{\partial}{\partial t} - i\mathbf{i}\nabla + \mathbf{j}m\right)\psi = -e(+i\mathbf{k}\phi + i\mathbf{i}\mathbf{A})\psi$$

We now apply a perturbation expansion to $\psi$, so that

$$\psi = \psi_0 + \psi_1 + \psi_2 + \ldots ,$$

with

$$\psi_0 = (i\mathbf{k}E + i\mathbf{i}\mathbf{p} + \mathbf{j}m)\, e^{-i(Et - \mathbf{p}\cdot\mathbf{r})}$$

as the solution of the unperturbed equation:

$$\left(-\mathbf{k}\frac{\partial}{\partial t} - i\mathbf{i}\nabla + \mathbf{j}m\right)\psi = 0 ,$$

which represents zeroth-order coupling, or a free fermion of momentum $\mathbf{p}$.

Using the perturbation expansion, we can write

$$\left(-\mathbf{k}\frac{\partial}{\partial t} - i\mathbf{i}\nabla + \mathbf{j}m\right)(\psi_0 + \psi_1 + \psi_2 + \ldots) = -e(\mathbf{k}\phi + i\mathbf{i}\mathbf{A})(\psi_0 + \psi_1 + \psi_2 + \ldots),$$

from which we can extract the first-order coupling as

$$\left(-\mathbf{k}\frac{\partial}{\partial t} - i\mathbf{i}\nabla + \mathbf{j}m\right)\psi_1 = -e(\mathbf{k}\phi + i\mathbf{i}\mathbf{A})\psi_0 .$$

If we expand $(\mathbf{k}\psi + i\mathbf{i}\mathbf{A})$ as a Fourier series, and sum over momentum $\mathbf{k}$, we obtain

$$(k\psi + ii\mathbf{A}) = \psi\,(k\psi(\mathbf{k}) + ii\mathbf{A}\,(\mathbf{k}))\,e^{i\mathbf{k}.\mathbf{r}},$$

so that

$$\left(-k\frac{\partial}{\partial t} - ii\nabla + jm\right)\psi_1 = -e\sum(k\phi(\mathbf{k}) + ii\mathbf{A}(\mathbf{k}))e^{i\mathbf{k}.\mathbf{r}}\psi_0$$

$$= -e\sum(k\phi(\mathbf{k}) + ii\mathbf{A}(\mathbf{k}))e^{i\mathbf{k}.\mathbf{r}}(ikE + i\mathbf{p} + jm)e^{-i(Et-\mathbf{p}.\mathbf{r})}$$

$$= -e\sum(k\phi(\mathbf{k}) + ii\mathbf{A}(\mathbf{k}))(ikE + i\mathbf{p} + jm)e^{-i(Et-(\mathbf{p}+\mathbf{k}).\mathbf{r})}$$

If we now expand $\psi_1$ as

$$\psi_1 = \sum v_1(E, \mathbf{p}+\mathbf{k})e^{-i(Et-(\mathbf{p}+\mathbf{k}).\mathbf{r})}$$

then

$$\sum\left(-k\frac{\partial}{\partial t} - ii\nabla + jm\right)v_1(E, \mathbf{p}+\mathbf{k})e^{-i(Et-(\mathbf{p}+\mathbf{k}).\mathbf{r})}$$

$$= -e\sum(k\phi(\mathbf{k}) + ii\mathbf{A}(\mathbf{k}))(ikE + i\mathbf{p} + jm)e^{-i(Et-(\mathbf{p}+\mathbf{k}).\mathbf{r})}$$

and

$$\sum(ikE + i(\mathbf{p}+\mathbf{k}) + jm)v_1(E, \mathbf{p}+\mathbf{k})e^{-i(Et-(\mathbf{p}+\mathbf{k}).\mathbf{r})}$$

$$= -e\sum(k\phi(\mathbf{k}) + ii\mathbf{A}(\mathbf{k}))(ikE + i\mathbf{p} + jm)e^{-i(Et-(\mathbf{p}+\mathbf{k}).\mathbf{r})}$$

and, equating individual terms,

$$(ikE + i(\mathbf{p}+\mathbf{k}) + jm)\,v_1(E, \mathbf{p}+\mathbf{k}) = -e\,(k\psi(\mathbf{k}) + ii\mathbf{A}(\mathbf{k}))(ikE + i\mathbf{p} + jm).$$

We can write this in the form

$$v_1(E, \mathbf{p}+\mathbf{k}) = -e[ikE + i(\mathbf{p}+\mathbf{k}) + ijm]^{-1}(k\psi(\mathbf{k}) + ii\mathbf{A}(\mathbf{k}))(ikE + i\mathbf{p} + jm)$$

which means that

$$\psi_1 = -e\sum[ikE + i(\mathbf{p}+\mathbf{k}) + jm]^{-1}(k\phi(\mathbf{k}) + ii\mathbf{A}(\mathbf{k}))(ikE + i\mathbf{p} + jm)e^{-i(Et-(\mathbf{p}+\mathbf{k}).\mathbf{r})}$$

which is the wavefunction for first-order coupling, with a fermion absorbing or emitting a photon of momentum **k**.

However, if we observe the process in the rest frame of the fermion and eliminate any *external* source of potential, then **k** = 0, and $(k\psi + ii\mathbf{A})$ reduces to the static value, $k\psi$, with $\psi$ as a self-potential. In this case, $\psi_1$ becomes

$$\psi_1 = -e[ikE + i\mathbf{p} + jm]^{-1}(k\phi)(ikE + i\mathbf{p} + jm)e^{-i(Et-\mathbf{p}.\mathbf{r})},$$

as the summation is no longer strictly required for a single order of the pure self-interaction. Since we can also write this as

$$\psi_1 = -e(ikE + i\mathbf{p} + jm)(ikE + i\mathbf{p} + jm)(k\phi)e^{-i(Et-\mathbf{p}.\mathbf{r})} , \qquad (44)$$

we see that $\psi_1 = 0$, for any fixed value of $\psi$. Clearly, this will also apply to higher orders of self-interaction. In other words, a *non-interacting* nilpotent fermion requires no renormalization as a result of its self-energy. The process could also be adapted for interacting particles, subject to external potentials, for here we can imagine redefining the $E$ and $\mathbf{p}$ operators to incorporate external potentials to make them 'internal', while simultaneously changing the structure of the phase factor to accommodate this. The change of phase factor would, of course, require a corresponding change in the amplitude, which could be taken as redetermining the value of the coupling constant, $e$, as required. Ultimately, however, (44) shows that it is the structure of ($ikE + i\mathbf{p} + jm$) as a nilpotent which seemingly eliminates the infinite self-interaction terms in the perturbation expansion at the same time as showing that they are merely an expression of the nature of the nilpotent vacuum as a reflection of the exactly supersymmetric nature of the original particle state.

**12 Cancellation of loops**

If the argument in section 11 is correct, then we should also be able to use the supersymmetric properties of the nilpotent operator to cancel fermion and boson loops directly. This is *exactly* what we would expect from a nilpotent system, where the total energy is zero, and one way of realising this would be to combine negative energy fermions with positive energy bosons. In the nilpotent formulation, every fermionic state has an intrinsic supersymmetric spin 1 bosonic vacuum partner with the same energy, momentum and mass. If we represent a spin ½ fermion by, say, ($\pm ikE \pm i\mathbf{p} + jm$), and a spin $-$½ fermion by ($\pm ikE \mp i\mathbf{p} + jm$), then each of these is unchanged by postmultiplication any number of times by the vacuum operator $k$ ($\pm ikE \pm i\mathbf{p} + jm$) or $k$ ($\pm ikE \mp i\mathbf{p} + jm$). However

($\pm ikE \pm i\mathbf{p} + jm$) $k$ ($\pm ikE \pm i\mathbf{p} + jm$) $k$ ($\pm ikE \pm i\mathbf{p} + jm$) $k$ ($\pm ikE \pm i\mathbf{p} + jm$) ...

and

($\pm ikE \mp i\mathbf{p} + jm$) $k$ ($\pm ikE \mp i\mathbf{p} + jm$) $k$ ($\pm ikE \mp i\mathbf{p} + jm$) $k$ ($\pm ikE \mp i\mathbf{p} + jm$) ...

are indistinguishable from

($\pm ikE \pm i\mathbf{p} + jm$) ($\mp ikE \pm i\mathbf{p} + jm$) ($\pm ikE \pm i\mathbf{p} + jm$) ($\mp ikE \pm i\mathbf{p} + jm$) ...

and

($\pm ikE \mp i\mathbf{p} + jm$) ($\mp ikE \mp i\mathbf{p} + jm$) ($\pm ikE \mp i\mathbf{p} + jm$) ($\mp ikE \mp i\mathbf{p} + jm$) ...

which alternate spin ½ and spin $-$½ fermions with spin 1 and spin $-$1 bosons. In effect the fermion generates its own vacuum boson partner, with the same $E$, $\mathbf{p}$ and $m$. Since the nilpotent structure is founded on zero totality, with the vacuum and fermion being in both zero superposition and zero combination, we may assume that this is an indication that the total energy made by positive boson and negative fermion loops is zero.

Now, the vacuum energy for a particle of mass *m* and spin *j* is given by [19]:

$$\frac{1}{2}(-1)^{2j}(2j+1)\int d^3k \sqrt{k^2+m_j^2} = \frac{1}{2}(-1)^{2j}(2j+1)\int d^3k \sqrt{k^2}\left(1+\frac{1}{2}\frac{m_j^2}{k^2}-\left(\frac{m_j^2}{k^2}\right)^2+...\right)$$

To remove the quartic, quadratic and logarithmic divergences, we need to ensure that

$$\sum_j (-1)^{2j}(2j+1) = 0 \tag{45}$$

$$\sum_j (-1)^{2j}(2j+1)m_j^2 = 0 \tag{46}$$

$$\sum_j (-1)^{2j}(2j+1)m_j^4 = 0 \tag{47}$$

Condition (45) requires equal numbers of fermionic and bosonic degrees of freedom. If we have $j = \pm \frac{1}{2}$ for the fermionic loops and $j = \pm 1$ for the bosonic loops, then

$$(-)^{2j}(2j+1)^{2j} = -2 \quad \text{for} \quad j = \frac{1}{2}$$
$$(-)^{2j}(2j+1)^{2j} = 3 \quad \text{for} \quad j = 1$$
$$(-)^{2j}(2j+1)^{2j} = 0 \quad \text{for} \quad j = -\frac{1}{2}$$
$$(-)^{2j}(2j+1)^{2j} = -1 \quad \text{for} \quad j = -1$$

giving a total of

$$\sum_j (-1)^{2j}(2j+1) = -2+3+0-1 = 0$$

as required.

Conditions (46) and (47) additionally require the fermions and bosons to have equal masses, which is true if the supersymmetry is intrinsic. Since all three conditions are fulfilled in the nilpotent formalism, it would appear that the intrinsic supersymmetry automatically removes the ultraviolet divergence.

In the case of a spin 0 boson (e.g. Higgs), we have a fundamental structure of either

$$(\pm ikE \pm i\mathbf{p} + jm)(\mp ikE \mp i\mathbf{p} + jm)$$

or

$$(\pm ikE \mp i\mathbf{p} + jm)(\mp ikE \pm i\mathbf{p} + jm)$$

with a combination of spin ½ and spin −½ fermions / antifermions (to which we can again apply vacuum operators). (The application of vacuum operators to the two partners in the combination would leave alternate creations of fermion and boson as before.) Since

$$(-)^{2j}(2j+1)^{2j} = 1 \quad \text{for} \quad j = 0$$

we can find a combination of spin ½ and spin 0, together with spin –½ and spin 0, which will lead to

$$\sum_j (-1)^{2j}(2j+1) = -2 + 1 + 0 + 1 = 0$$

again as required, and, with *m* common to fermions and bosons, also fulfilling conditions (46) and (47). It would appear from this argument that the divergence is again removed and, in particular, that there is no reason to expect a hierarchy problem for the Higgs boson.

One further physical problem to which this may relate is the matter / antimatter asymmetry between fermions and antifermions. This is a long-standing problem for which answers have been generally sought in cosmology. However, the asymmetry could in fact be generic. It could reflect that we have defined two vector spaces, characterised in one representation by positive and negative energies. If, however, we see fermions (with *E*) as being the characteristic particles defining real (observable) space, then we could see antifermions (with –*E*) as necessarily being the characteristic particles defining vacuum space. In that case, we would not expect a symmetry between the two particle types in either of the spaces.

## 13 Propagators

The definition of a physical singularity as emerging from the combination of two dual vector spaces and the *zitterbewegung* that this generates through the switching between them (which is equivalent to the switching between +*iE* and –*iE*) can be seen as the origin of the 'pole' or singularity that appears in the particle propagator in the conventional Feynman formalism. This appears in the nilpotent formalism and is a classic sign of the action of vacuum, generally taken as the point of 'switchover' between fermion and antifermion states, paralleling the dual vector spaces of the theory through complex analysis. However, in nilpotent theory, the pole is no longer a 'naked' singularity, causing an infinite divergence, but one accommodated within the dual spaces on which the theory is founded. The nilpotent formalism incorporates the pole automatically without divergence because of its direct inclusion of vacuum states. Conventional theory assumes that a fermion propagator takes the form

$$S_F(p) = \frac{1}{\not{p} - m} = \frac{\not{p} + m}{p^2 - m^2},$$

where $\not{p}$ represents $\gamma^\mu \partial_\mu$, or its eigenvalue, and that there is a singularity or 'pole' ($p_0$) where $p^2 - m^2 = 0$, the 'pole' being the origin of positron states. On either side of the pole there are positive energy states moving forwards in time, and negative energy states moving backwards in time, the terms ($\not{p} + m$) and ($-\not{p} + m$) being used to project out, respectively, the positive and negative energy states. The normal solution is to add an infinitesimal term $i\varepsilon$ to $p^2 - m^2$, so that $iS_F(p)$ becomes

$$iS_F(p) = \frac{i(\not{p} + m)}{p^2 - m^2 + i\varepsilon} = \frac{(\not{p} + m)}{2p_0}\left(\frac{1}{p_0 - \sqrt{p^2 + m^2} + i\varepsilon} + \frac{1}{p_0 + \sqrt{p^2 + m^2} - i\varepsilon}\right)$$

and take a contour integral over the complex variable to give the solution

$$S_F(x-x') = \int d^3p \frac{1}{(2\pi)^3} \frac{m}{2E}\left[-i\theta(t-t')\sum_{r=1}^{2}\Psi(x)\overline{\Psi}(x') + i\theta(t'-t)\sum_{r=3}^{4}\Psi(x)\overline{\Psi}(x')\right]$$

with summations over the up and down spin states.

This mathematical subterfuge is unnecessary in the nilpotent formalism because the denominator of the propagator term is always a nonzero scalar. We write

$$S_F(p) = \frac{1}{(\pm ikE \pm i\mathbf{p} + jm)},$$

and choose our usual interpretation of the reciprocal of a nilpotent to give:

$$\frac{1}{(\pm ikE \pm i\mathbf{p} + jm)} = \frac{(\pm ikE \mp i\mathbf{p} - jm)}{(\pm ikE \pm i\mathbf{p} + jm)(\pm ikE \mp i\mathbf{p} - jm)} = \frac{(\pm ikE \mp i\mathbf{p} - jm)}{4(E^2 + p^2 + m^2)},$$

which is finite at all values. The integral is now simply

$$S_F(x-x') = \int d^3p \frac{1}{(2\pi)^3} \frac{m}{2E} \theta(t-t')\,\Psi(x)\,\overline{\Psi}(x'),$$

in which $\Psi(x)$ is the usual

$$\Psi(x) = (\pm ikE \pm i\mathbf{p} + jm)\exp(ipx),$$

with the phase factor written as a 4-vector, and the adjoint term becomes

$$\overline{\Psi}(x') = (\pm ikE \mp i\mathbf{p} + jm)(i\mathbf{k})\exp(-ipx').$$

Since the nilpotent formalism comes as a complete package with a single phase term, automatic second quantization, and the negative energy states matched with reverse time states, there is no averaging over spin states or separation of positive and negative energy states on opposite sides of a pole. The particle structure is itself the singularity. There is no division between the particle and antiparticle because the two come as a single unit incorporating real space and vacuum space on an equal footing.

The fermion propagator can also be used to define boson propagators. In conventional theory, we derive the boson propagator (48) directly from the Klein-Gordon equation, while recognizing that its mathematical form depends on the choice of gauge:

$$\Delta_F(x-x') = \frac{\not{p} + m}{p^2 - m^2}. \tag{48}$$

This is because the Klein-Gordon operator

$$\left(\gamma^0 \frac{\partial}{\partial t} + \boldsymbol{\gamma}.\nabla + im\right)\left(\gamma^0 \frac{\partial}{\partial t} + \boldsymbol{\gamma}.\nabla - im\right) = \left(\frac{\partial^2}{\partial t^2} - \nabla^2 + m^2\right)$$

is the only scalar product which can emerge from a differential operator defined as in (1). The Klein-Gordon equation, however, is not specific to boson states or an identifier of them. It merely defines a universal zero condition which is true for all states, whether bosonic or fermionic. And, the propagator in (48) does not correspond to any known bosonic state. Instead, we have *three* boson propagators.

Spin 1: $\quad \Delta_F(x - x') = \dfrac{1}{(\pm ikE \pm i\mathbf{p} + jm)(\mp ikE \pm i\mathbf{p} + jm)}$,

Spin 0: $\quad \Delta_F(x - x') = \dfrac{1}{(\pm ikE \pm i\mathbf{p} + jm)(\mp ikE \mp i\mathbf{p} + jm)}$,

Paired Fermion: $\quad \Delta_F(x - x') = \dfrac{1}{(\pm ikE \pm i\mathbf{p} + jm)(\pm ikE \mp i\mathbf{p} + jm)}$.

Where the spin 1 bosons are massless (as in QED), we will have expressions like:

$$\Delta_F(x - x') = \frac{1}{(\pm ikE \pm i\mathbf{p})(\mp ikE \pm i\mathbf{p})}. \tag{49}$$

Clearly, the relationship of the fermion and boson propagators is of the form

$$S_F(x - x') = (i\gamma^\mu \partial_\mu + m)\, \Delta_F(x - x'),$$

or, in our notation,

$$S_F(x - x') = (\pm ikE \pm i\mathbf{p} + jm)\, \Delta_F(x - x').$$

which is exactly the same relationship as is defined between fermion and boson in the nilpotent formalism.

Now, using

$$iS_F(p) = \frac{1}{2p_0}\left(\frac{1}{p_0 - \sqrt{p^2 + m^2} + i\varepsilon} + \frac{1}{p_0 + \sqrt{p^2 + m^2} - i\varepsilon}\right),$$

which is the same as the conventional fermion propagator up to a factor ($\not{p} + m$), we can perform a contour integral which is similar to that for the fermion to produce

$$i\Delta_F(x - x') = \int d^3p\, \frac{1}{(2\pi)^3}\, \frac{1}{2\omega}\, \theta(t - t')\, \phi(x)\phi^*(x').$$

Here, $\omega$ takes the place of $E / m$, while $\phi(x)$ and $\phi(x')$ are now scalar wavefunctions. However, in our notation, they will be scalar products of ($\pm$ *ikE* $\pm$ ***i*p** + ***j**m*) exp (*ipx*) and ($\mp$ *ikE* $\pm$ ***i*p** + ***j**m*) exp (*ipx'*) and $\phi(x)\phi^*(x')$ reduces to a product of a scalar term, which can be removed by normalization, and exp $ip(x - x')$.

In off-mass-shell conditions, where $E^2 \neq p^2 + m^2$, poles in the propagator are a mathematical, rather than physical, problem, and removed by the use of $i\varepsilon$ and the contour integral, which is *ad hoc* but effective. However, in the specific case of massless bosons, conventional theory cannot prevent 'infrared' divergencies appearing in (48) when such bosons are emitted from an initial or final stage which is on the mass shell. Such divergencies, however, do not occur where there is no naked pole, as in (49). The definition of the boson propagator as (49), rather than (48), not only shows that one of the principal divergences in quantum electrodynamics is, as the procedure used to remove it would suggest, merely an artefact of the mathematical structure we have imposed, and not of a fundamentally physical nature, but also suggests that the formalism which removes it is a more exact representation of the fundamental physics. Ultimately, this is because it allows an exact representation of the vacuum simultaneously with the fermionic state.

## 14 Weak interactions

One of the most important aspects of the nilpotent structure (with its pseudoscalar, vector and scalar components) is that it *already incorporates the fundamental interactions*. Simply defining a nilpotent fermion by this mathematical formalism means that it is *necessarily* acting according to some or all of these interactions. They arise solely from its internal structure. Coulomb terms, for example, are simply the result of spherical symmetry of point sources. Since this Coulomb interaction is purely an expression of the magnitude of a scalar phase, all the terms in the nilpotent contribute, but only one, the passive (scalar) mass term, contributes to nothing else. An interaction with this precise property may therefore be defined, and it is the one we define as the *electric* interaction. At the same time, the strong interaction, with its characteristic linear potential, can be represented as we have seen, by the vector properties of the **p** term.

However, yet another interaction seems to be required by the *spinor* structure of the nilpotent operator, and the associated phenomenon of *zitterbewegung*. While the co-existence of two spin states is, in some sense, real, and is accounted for by the presence of mass, the co-existence of two energy states is only meaningful in the context of the simultaneous existence of fermion and vacuum. While the transitions between the two energy states may be virtual, in this sense, the *zitterbewegung* would seem to require the production of an intermediate bosonic state at a vertex where one fermionic state is annihilated and another is created to replace by it. This behaviour is, of course, characteristic of the weak interaction, and, in this sense, we can say that the weak interaction, like the electric and strong interactions, is built into the structure of the nilpotent operator.

The weak interaction is clearly related to the nature of the pseudoscalar *iE* operator, whose sign uniquely determines the helicity of a weakly interacting particle, or more specifically its weakly interacting component. It also has a unique feature, in that its fermionic source cannot be separated from its vacuum partner. A fermion or antifermion cannot be created or annihilated, even with an antifermionic or fermionic partner, unless its vacuum is simultaneously annihilated or created. In this sense, the weak source has a manifestly dipolar nature, whose immediate manifestation is the fermion's ½-integral spin. It is the most direct evidence we have of the duality of the vector space structure which underlies quantum physics. This, then, leads to the question of whether we can derive an analytic expression from the nilpotent operator which will explain the special characteristics of this force. To answer this, it will be convenient to answer a more general question: what nilpotent solutions are available for an operator including a Coulomb potential together with any other potential which is a function of distance from a point source with spherical symmetry, other than the linear potential characteristic of the strong interaction?

We will assume that the nilpotent operator takes a form such as

$$\left( k\left( E - \frac{A}{r} - Cr^n \right) + i\left( \frac{\partial}{\partial r} + \frac{1}{r} \pm i\frac{j+½}{r} \right) + ijm \right) . \tag{50}$$

where $n$ is an integer greater than 1 or less than $-1$, and, as usual, look for a phase factor which will make the amplitude nilpotent. Again, we will work from the basis of the Coulomb solution, with the additional information that polynomial potential terms which are multiples of $r^n$ require the incorporation into the exponential of terms which are multiples of $r^{n+1}$. So, extending our work on the Coulomb solution, we may suppose that the phase factor is of the form:

$$\phi = \exp(-ar - br^{n+1}) r^\gamma \sum_{\nu=0} a_\nu r^\nu$$

Applying the operator and squaring to zero, with a termination in the series, we obtain

$$4\left( E - \frac{A}{r} - Cr^n \right)^2 = -2\left( -a + (n+1)br^n + \frac{\gamma}{r} + \frac{\nu}{r} + \frac{1}{r} + i\frac{j+½}{r} \right)^2$$
$$-2\left( -a + (n+1)br^n + \frac{\gamma}{r} + \frac{\nu}{r} + \frac{1}{r} - i\frac{j+½}{r} \right)^2$$

Equating constant terms, we find
$$a = \sqrt{m^2 - E^2} \tag{51}$$

Equating terms in $r^{2n}$, with $\nu = 0$:
$$C^2 = -(n+1)^2 b^2$$
$$b = \pm \frac{iC}{(n+1)} .$$

Equating coefficients of $r$, where $\nu = 0$:

$$AC = -(n+1)b(1+\gamma),$$
$$(1+\gamma) = \pm iA .$$

Equating coefficients of 1 / $r^2$ and coefficients of 1 / $r$, for a power series terminating in $v = n'$, we obtain

$$A^2 = -(i\gamma n')^2 + (j + \tfrac{1}{2})^2 \qquad (52)$$

and

$$-EA = a(i\gamma + n'). \qquad (53)$$

Combining (51), (52) and (53) produces:

$$\left(\frac{m^2 - E^2}{E^2}\right)(1 + \gamma + n')^2 = -(1 + \gamma + n')^2 + (j + \tfrac{1}{2})^2$$

$$E = -\frac{m}{j + \tfrac{1}{2}}(\pm iA + n'). \qquad (54)$$

Equation (54) has the form of a harmonic oscillator, with evenly spaced energy levels deriving from integral values of $n'$. It does not immediately suggest the value for the term $iA$, but, if we make the additional assumption that $A$, the phase term required for spherical symmetry, has some connection with the random directionality of the fermion spin, we might assign to it a half-unit value ($\pm \tfrac{1}{2} i$), or ($\pm \tfrac{1}{2} i\hbar c$), using explicit values for the constants, and obtain the complete formula for the fermionic simple harmonic oscillator:

$$E = -\frac{m}{j + \tfrac{1}{2}}(\tfrac{1}{2} + n'). \qquad (55)$$

Now, the dimensions of $A$ are those of charge ($q$) squared or interaction energy × range, and an $A$ numerically equal to $\pm \tfrac{1}{2} \hbar c$ would be exactly that required by the uncertainty principle, allowing the value of the range of an interaction mediated by the $Z$ boson to be calculated as $\hbar / 2M_Z c = 2.166 \times 10^{-18}$ m, as observed. The $\tfrac{1}{2} \hbar c$ term is also significant in the expressions for zero-point energy and *zitterbewegung*, which connect with both spin and the uncertainty principle. Interpreting the *zitterbewegung* as a dipolar switching between fermion and vacuum antifermion states, we can describe this in terms of a weak dipole moment $(\hbar c / 2)^{3/2} / M_Z c^2$, of magnitude $8.965 \times 10^{-18}$ $e$ m ($1.44 \times 10^{-36}$ Cm). Because of the specific appearance of the $\tfrac{1}{2} \hbar c$ term for spin ($s$) in $\mu = gqs / 2m$, an identical expression can additionally be used to define a weak magnetic moment, of order $4.64 \times 10^{-5}$ × the magnetic moment of the electron. The existence of such a dipole moment would make the spin $\tfrac{1}{2}$ term an expression of the dipolarity of the weak vacuum, and a physical representation of the weak interaction as a link between the fermion and vacuum, or between real space and vacuum space. The possible appearance of an imaginary factor $i$ in $A$ is interesting in relation to the requirement of a complex potential or vacuum for *CP* violation in the pure weak interaction.

Whatever assumptions we make about $A$, equation (54) demonstrates that the additional potential of the form $Cr^n$, where $n$ is an integer greater than 1 or less than –1, has the effect of creating a harmonic oscillator solution for the nilpotent operator, irrespective of the value of $n$, and, in fact, we can show that any polynomial sum of potentials of this form will produce the same result. Such potentials emerge from any system in which there is complexity, aggregation, or a multiplicity of sources, even if the individual sources have Coulomb or linear potentials. In the case of a dipolar weak sources, there will be a minimum extra term of the form $Cr^{-3}$, and so we can say that (55) provides the correct characteristics for the weak interaction from the kind of potential that weak sources must necessarily produce. In addition, because this solution is exclusive for distance related potentials of the form $Cr^n$, except where $r = 1$ or $-1$, we have also, in effect, shown that a fermion interaction specified in relation to a spherically symmetric point source has only three physical manifestations, and that these are the ones associated with the electric (or other pure Coulomb), strong and weak interactions.

## 15 Mass generation

In string theory, mass is generated by the vibrations of the strings, which replace point particles. However, this mass-generating mechanism is already incorporated in the point particle concept (as the Lamb shift makes clear), and relates to the Berry phase and *zitterbewegung*. It comes from the dual vector spaces needed to define a point particle, because the duality ensures that *zitterbwegung* (and hence vacuum fluctuation, the Lamb shift, etc.) is the origin of fermionic mass, and it *requires* a pole or singularity. In the nilpotent theory, rest mass always comes from defining a singularity through a double vector space. The very act of defining a point particle is also the same as ensuring that it undergoes vacuum fluctuations, or equivalent, and therefore generates mass. Again connecting with string theory, it is the same duality as that between gravity and gauge theory or between the local and nonlocal.

The nilpotent operator ($\pm ikE \pm i\mathbf{p} + jm$) can, in fact, be regarded as a 10-D object (embedded in Hilbert space): 5 for $iE$, $\mathbf{p}$, $m$ and 5 for $k$, $i$, $j$; and six of these (all but $iE$ and $\mathbf{p}$) are compactified. The fact that they are not all *spatial* dimensions is irrelevant if string theory produces its extra dimensions below the Planck length. A classic prescription for a perfect string theory is one in which 'self-duality in phase space determines vacuum selection'. The nilpotent certainly fulfils this criterion and it is also a mass-shell system and incorporates the right groups. Though we have no need for a model-dependent theory to incorporate the interactions, it is important to be able to satisfy all the conditions that appeared to make string theory, *or a more fundamental abstract theory, of which the model-dependent theories are approximations*, seemingly necessary. It is significant that the nilpotent formalism achieves this through solving the problem of vacuum.

The Higgs mechanism has provided a process for mass generation in the Standard Model and this has been discussed, in terms of the nilpotent formalism, in section 8. However, as is well known, the mechanism provides no method of generating the actual mass of the Higgs boson. Schlücker [20] has produced a list of 78 different predictions, covering the entire range from 100 GeV to 500 GeV, with a few higher ones. As so many additional criteria have been applied, at this stage one guess is as good as another. My own guess, calculated from a vacuum argument, provides an interestingly exact value of 181.5 GeV (from 2596 $m_e / \alpha$), but can claim no special authority [1]. One of the interesting aspects of this result is that it appears to be close to the sum of all the fermion masses (as well as to being near to $2M_Z$), and so could suggest a 'partitioning' of the vacuum by Higgs coupling to provide the masses of individual fermions.

The closest result in Schlücker's list is that of Namsrai [21], who predicts a value of $2M_Z$, on the basis that, with space-time curvature, a self-referential nonlinear field will generate soliton-like solutions and 'wave'-like structures that will allow particles to acquire mass by Yukawa coupling. Namsrai's argument is interesting, especially in the fact that the factor 2 is generated (as elsewhere) by curvature. His Higgs particle, however, 'is essentially nonlocal and spreads out over the whole of space; its propagator has no pole in momentum space. It is perhaps difficult' [and probably almost impossible] 'to detect'. In the theory presented here, however, rest mass can only arises locally and through *zitterbewegung*; it requires a real singularity, essentially generating a pole in the kind of spin 0 propagator discussed in section 13. A locally-described Higgs boson of mass 181.5 GeV would probably decay observably via two *Z* bosons.

**16 Conclusion: some fundamental dualities**

We have seen that, in the nilpotent formalism, the operator and wavefunction are dual, but there is also duality at many other levels in the structure. A related dualism between fermion and vacuum originates in the idea that, by defining a fermion state, we are also defining a fundamental singularity. In principle, we can only define something as a singularity if we define *everything else* at the same time as excluded from the definition. To define a singularity we are forced to use a dualistic structure by simultaneously defining what is not singular. If we can view the fermion as a singularity with connections leading out to the rest of the universe, we can see the vacuum as a kind of 'inverse singularity', with connections from the rest of the universe leading into the singularity that constitutes the fermion state. Pauli exclusion can then be reinterpreted as saying that no fermion state can share its vacuum with any other.

The duality described here ensures that vacuum is not something separated from the fermion. It is an intrinsic component of its definition, and of the spinor structure needed to define the fermion as a singular state. It is the reason why the fermion has half-integral spin – we can only define it by simultaneously splitting the universe into two halves which are mirror images of each other. The duality manifests itself physically in the phenomenon of *zitterbewegung*. Using either operator or amplitude, we define ($\pm i k E \pm i \mathbf{p} + j m$) as a 4-spinor, with 4 terms (each of which is nilpotent) arranged as a column / row vector.

In the convention used here, the 'real' state (the one subject to physical observation) is determined by the signs of *E* and **p** in the first term. The other three states are like three 'dimensions' of vacuum, the states into which the real term could transform by respective *P*, *T* or *C* transformations. The duality ensures that fermion and vacuum occupy separate 3-dimensional 'spaces', which are combined in the $\gamma$ algebra defining the singularity state. It can be shown that these 'spaces', though seemingly different, are truly dual, each containing the same information, and that this duality manifests itself directly in many physical forms.

One of these can be seen when we apply the conventional explanation of Pauli exclusion, that the wavefunctions of fermions are antisymmetric (which reduces purely to information about the **p** vector) and compare it to that defined by nilpotency. We can imagine, in the latter case, that the *iE*, **p** and *m* terms are represented on orthogonal axes, and that the direction of the resultant 'vector' is unique for each state. In principle, we have two mappings for Pauli exclusion, on $\sigma^1$, $\sigma^2$, $\sigma^3$ and on $\Sigma^1$, $\Sigma^2$, $\Sigma^3$, showing that these two sets of 3-dimensional coordinates are dual. That this is not merely a mathematical trick is shown by the fact that both sets of coordinates yield information about the same physical quantity: angular momentum. For full specification, angular momentum requires three separate pieces of information – magnitude, direction and handedness – and this is provided when *iE*, **p** and *m* are combined. It is also provided when we use all the information incorporated in the **p** vector alone.

Another example of the duality occurs with the derivation of spin ½ for fermions. The standard approach, as in section 5, is to take the commutator of the spin pseudovector **σ = –1** and the Hamiltonian, of which the only anticommutating component is **p**. The resulting factor 2, which leads to half-integral spin, then comes from the anticommuting aspects of the components of **p**. Here, spin ½ derives totally from the multivariate (Pauli matrix) nature of **p** and has nothing to do with relativity at all. This is further demonstrated by the fact that the anomalous magnetic moment, based on spin ½, can be derived from the nonrelativistic Schrödinger equation [22]. However, historically, spin ½ was first explained in terms of the Thomas precession, which is, of course, a relativistic correction. In other words, we can derive spin ½ using either the (multivariate) vector properties of space (using $\sigma^1$, $\sigma^2$, $\sigma^3$) *or* the relativistic connection between space and time (using $\Sigma^1$, $\Sigma^2$, $\Sigma^3$); the 3-dimensional 'spaces' involved are totally dual.

The same can be said for the velocity addition law in special relativity, which can be derived using either two dimensions of space (which generates the $\sigma^1$, $\sigma^2$, $\sigma^3$ structure of Euclidean space) or one of space relativistically connected with one of time (which generates the $\Sigma^1$, $\Sigma^2$, $\Sigma^3$ connection between space, time and proper time). Another example where this occurs is the holographic principle (which is completely defined for the fermionic case by the nilpotent structure) where the bounding 'area' can be defined either by two spatial coordinates or one of space and one of time [23, 24]. (The fact that the holographic principle has featured notably in the discussion of gravitational singularities may well suggest that these can also be defined in terms of a dual space structure.)

All these examples are characteristic cases in which the two vector spaces required to define fermion structure as a singularity are completely dual, even though the symmetry of one is preserved while that of the other is broken. It would seem that one condition is necessary to define the opposite in the other, and that, in this way, the opposing conditions ultimately provide the same information, in the same way as localised fermion state and nonlocalised vacuum, or operator acting on phase factor and amplitude. *Zitterbewegung* has been interpreted as a switching between a fermion state and its vacuum; it is also an expression of the duality between the 'real' space of $\sigma^1$, $\sigma^2$, $\sigma^3$ and the 'vacuum space' of $\Sigma^1$, $\Sigma^2$, $\Sigma^3$, neither of which is privileged. Both give an equally correct description of the state and must be simultaneously valid, even though we can only observe one at any given moment, and even the choice of broken / unbroken rotation symmetries between the components can be reversed by switching the space of observation from 'real' to 'vacuum' space.

It is significant that the terms in (18) can be rearranged so that the nature of the real state is any of them. The other three terms then rearrange themselves by automatic sign variation as the corresponding vacuum states. Thus, we may define the 'real' structures:

$$(\pm ikE \pm i\mathbf{p} + jm) \qquad \text{fermion spin up}$$
$$(\pm ikE \mp i\mathbf{p} + jm) \qquad \text{fermion spin down}$$
$$(\mp ikE \pm i\mathbf{p} + jm) \qquad \text{antifermion spin down}$$
$$(\mp ikE \mp i\mathbf{p} + jm) \qquad \text{antifermion spin up} \qquad (56)$$

If a fermion could combine with its own vacuum, it would annihilate automatically, but this is, of course, impossible because this vacuum represents the entire universe outside of the fermion; however, we can imagine it combining with a *component* of vacuum, or the fermion state equivalent to this. (A 4-D combination of this kind seems to be the minimum structure to describe change in a 3-D system.) Apart from annihilation with the rest of the universe, we can then identify 3 further scalar products between fermionic states with the same values of $E$ and $\mathbf{p}$, though this time with different signs. All of these, as we have seen, appear to have a physical realisation:

$$(\pm ikE \pm i\mathbf{p} + jm)(\pm ikE \pm i\mathbf{p} + jm) \qquad \text{universal zero totality}$$
$$(\pm ikE \pm i\mathbf{p} + jm)(\pm ikE \mp i\mathbf{p} + jm) \qquad \text{fermion-fermion combination}$$
$$(\pm ikE \pm i\mathbf{p} + jm)(\mp ikE \pm i\mathbf{p} + jm) \qquad \text{spin 1 boson}$$
$$(\pm ikE \pm i\mathbf{p} + jm)(\mp ikE \mp i\mathbf{p} + jm) \qquad \text{spin 0 boson} \qquad (57)$$

These combinations between two fermion / antifermion states may also be seen as the vertex of an interaction or the combination of one fermion state with the vacuum of another. The spin 1 and spin 0 bosons states represented in (57) have all the properties of such states observed in physics experiments. The states represented here may be seen as simultaneous realisations of the two vector spaces involved in the creation of the fermionic state, though at the expense of making only one component of the vacuum space well defined, just as only one component of angular momentum is well defined in real space. They are effectively combinations of a fermion with its own vacuum, subjected to a *P*, *T* or *C* transformation. It is in this context also that we can use the (discrete) *idempotent* properties inherent in the nilpotent quantity $(\pm ikE \pm i\mathbf{p} + jm)$ to provide such vacuum states.

The intrinsically dualistic nature of the fermion is most readily apparent when it is described by the self-dual nilpotent form of quantum mechanics, which is founded on the commutative combination of two vector spaces, each of which is exactly dual to the other. From this initial duality, many others emerge, for example, those between fermion and vacuum, fermion and vacuum boson, operator and amplitude, nilpotent and idempotent, broken and unbroken symmetries. These dualities allow the same mathematical structures (or the same structures but for sign changes) to describe apparently dissimilar objects, and so explain how the creation of a fermionic singularity effectively splits the universe into two halves that are mathematically and physically, if not observationally, equivalent.